\documentclass[12pt,eqsecnum]{article}             

\usepackage{graphics}
\usepackage{latexsym}
\usepackage{epsfig}
\usepackage{amsbsy}
\usepackage{amsmath}

\begin{document}

\title{Photon coincidence spectroscopy for	\\
	two--atom cavity quantum electrodynamics}

\author{L.\ Horvath
	and B.\ C.\ Sanders}

%%\address{
%{\em Department of Physics, Macquarie University, %institute<-->address
%Sydney, New South Wales 2109, Australia}
%
\date{Received: 2000 / Revised version: } 
\maketitle

\abstract{   
We show that photon coincidence spectroscopy can provide an unambiguous
signature of two atoms simultaneously interacting with a quantised 
cavity field mode.
We also
show that the single--atom Jaynes--Cummings model can be probed
effectively via photon coincidence spectroscopy, even with
deleterious contributions to the signal from two--atom events. 
In addition, we have explicitly 
solved the eigenvectors and eigenvalues of two two--level
atoms coupled to a quantised cavity mode for differing coupling strengths.

%\PACS{  
%      {PACS-key}{42.50.Ct} \and
%      {PACS-key}{42.50.D} }% end of PACS codes
}%end of abstract

%\maketitle %comment out
%

%\setlength{\textwidth}{11cm}
%\onecolumn

\section{Introduction}
\label{intro}

Photon coincidence spectroscopy (PCS) has been introduced as a tool for
probing the spectrum of the Jaynes--Cummings (JC) model corresponding
to a single two--level atom (2LA)
coupled to one quantised field mode of an optical resonator.
Such a system is realised via
atomic beam cavity quantum electrodynamics (CQED)
\cite{Rai89,Zhu90,Tho92,Tur95,Bru96,Car96,San97,Hor99,Hor00,Hor00a}.
In two--photon coincidence spectroscopy (2PCS),
a sparse atomic beam traversing the optical cavity is driven off--axis
by a bichromatic driving field, and two photons in the output field of
the cavity are deemed to be coincident if they arrive at the
photodetectors within some short time interval~\cite{Car96,Hor00}.
During the 2LA's passage through the cavity, the electric dipole
coupling strength~$g(\mbox{\boldmath{$r$}})$ varies with its 
position~$\mbox{\boldmath{$r$}}$, and the coupling strength peaks at an
antinode along the cavity's longitudinal axis. The 2LAs are assumed
to be sufficiently slowly moving so that the system can be treated as 
a JC stationary system with a coupling strength $g$ varying
according to the  
probability distribution~$P(g)$.
Thus, atomic motion is responsible for an effective
inhomogeneous broadening of the spectrum for the coupled atom--cavity
system. However, 
the bichromatic driving field selectively excites a subensemble of
coupled atom--cavity systems (with particular values of~$g$)
to the second couplet of the JC
ladder.  The JC spectrum is then probed via two--photon
coincidence spectroscopy by measuring the rate of two--photon
coincidence events, which is referred to as the
two--photon coincidence rate (2PCR)~\cite{Car96,Hor00}.

A bichromatic driving field has been used in 
optical CQED~\cite{Tur95}, but photon coincidence measurements
are yet to be implemented in experiments.  More recently experiments
have succeeded in controlling, to some extent, single--atom motion in
the cavity~\cite{Pin00}, but complete confinement of a single atom in
a cavity~\cite{Doh01,Par00}
with~$P(g)\propto\delta(g-g_0)$, i.e.\  a
`frozen' atom, is yet to be achieved.  
2PCS is designed specifically to overcome
inhomogeneous broadening due to atomic motion, 
thereby permitting the nonlinear component of the 
JC spectrum (specifically the second couplet of the JC ladder and
its two photon de--excitation events) to be optically probed.

2PCS has been shown to be effective for overcoming the challenges of
inhomogeneous broadening due to atomic motion. However, a second
challenge is atom number fluctuations. As the atoms arrive at the
cavity randomly in an atomic beam, the number of atoms in the cavity
changes: most of the time no atom is in the cavity; some of the time
there is one atom; less frequently there are two or more atoms in the
cavity.

In quantum trajectory simulations of 2PCS, multi--atom effects 
have been included 
in the simulations~\cite{Car96}, and the nonlinear regime of the 
JC spectrum may be observed even in the presence of multi--atom
contributions to the photon statistics.  Of course a sparse beam of 2LAs is 
required to ensure that multi--atom effects do not overwhelm the photon
statistics. Rigorous bounds on the mean atom number in CQED have
been established \cite{Car99}.  Here we investigate in detail
the contributions
of multi--atom effects to the extraction of an unambiguous signature of 
the JC spectrum.
We show additional peak structure that arises in the two--photon coincidence
spectrum due to the two--2LA contributions, but we also show that these effects
can be negligible for a sparse beam.  An effect of multi--atom
contributions to the 2PCR 
is to raise the background 2PCR,
thereby making 2PCS only marginally more difficult than if two--atom effects
could be avoided altogether.

Furthermore,
we consider the system of two 2LAs 
coupled to a single cavity field mode.  For the sparse atomic beam passing
through the cavity, we show that an off--axis bichromatic driving field
can be used to observe the spectrum of the coupled system consisting
of two 2LAs and one quantised cavity field mode.  By choosing the
two frequencies of the bichromatic driving field judiciously, one
may extract the spectrum for various choices of coupling strengths~$g_1$
and~$g_2$ for atoms~$1$ and~$2$, respectively.  The case $g_1=g_2$ 
corresponds to probing the two--atom version of the
Tavis--Cummings model~\cite{Tav69}.

We develop the Hamiltonian for multi--atom systems in section~\ref{sec:math}.
In section~\ref{sec:twoatomspec}, we develop
2PCS as a tool for probing the spectrum of two atoms coupled to a single 
quantised cavity field mode, and we conclude in section~\ref{sec:conclusions}.

\section{Mathematical background}
\label{sec:math}

In the typical CQED experiment, which tests for quantum field effects, 
an atomic
beam is directed through a cavity~\cite{Tur95,Fos00}
and is excited by a driving field during the passage
of the atom.  In contrast to
microwave cavity quantum electrodynamics~\cite{Bru96,Scu96},
where the atomic lifetime is large compared to the passage time through the
cavity, the lifetime in an optical system is quite short, hence the need for
optically driving the system during the passage of the atom through 
the cavity.  Whereas
current experiments employ an on--axis driving field (the laser field
is directed into the cavity mode through one mirror), 2PCS is dramatically
improved for an off--axis driving field (the atoms are driven directly
during passage through the cavity by a bichromatic driving field),
and photon coincidences are detected in the cavity output field.

\subsection{The Hamiltonian}
\label{subsec:H}

The Hamiltonian for the cavity field is given by
\begin{equation}
\label{H0}
H^{(0)} = \hbar \omega a^\dagger a~,
\end{equation}
for~$n=a^{\dag}a$ the photon number operator, and 
$\omega$ is the angular frequency of the cavity mode
(assumed to be equal to the transition frequency for the
2LA).  If there are $N$ 2LAs in the cavity,
the Hamiltonian for the $m^{\rm th}$ 2LA is 
generalised from~$H^{(0)}$ of equation~(\ref{H0}) to yield
\begin{eqnarray}
\label{H:terms}
H^{(m)}(g_m)&=&\hbar\omega\sigma^{(m)}_z + i \hbar \left( 
	g_m 
	a^{\dagger} \sigma^{(m)}_- 
	- g^*_m  
	a \sigma^{(m)}_+\right), 	
\end{eqnarray}
with~$g_m\equiv g(\mbox{\boldmath{$r$}}_m)$ the coupling strength 
between the $m^{\rm th}$ atom and
the cavity field. 
Although the atom is moving, we assume that the motion is slow so that
time--dependence of the Hamiltonian can be ignored. Atomic motion is then
accounted for by allowing~$g_m$ to be a random variable~\cite{Car96,Hor00}.
Taking~$g_m = \vert g_m 
\vert \exp(i \theta_m)$, and 
applying a unitary 
transformation~$U^{(m)}(g_m)=\exp(i \theta_m \sigma_z^{(m)})$ to 
equation~(\ref{H:terms}), 
the Hamiltonian simplifies to
\begin{eqnarray}
\label{H:trans}
[U^{(m)}]^{\dag} (g_m)
H^{(m)}(g_m)U^{(m)}(g_m) = H^{(m)}(\vert g_m \vert).	
\end{eqnarray}
Thus, we use the simpler Hamiltonian~(\ref{H:trans}) for
treating the~$m^{\rm th}$ atom and assume that~$g_m$ is real
and positive.
The 2PCR is a function of the two--photon correlation
\begin{equation}
\label{eq:corr}
w^{(2)}(t)=\langle : n(t_0) n(t_0+t) : \rangle 
\end{equation}
for an unimportant `initial' time~$t_0$, and `$::$' imposes 
normal ordering.
The time interval for coincidences is assumed to be short so that
the 2PCR is approximately proportional to~\cite{Car96,San97,Hor00}
\begin{equation}
\label{eq:twocorr}
w^{(2)}=\langle :n^2 : \rangle.
\end{equation}

The atomic beam is slowly moving, and the time dependence of~$g_m$ can 
be ignored by instead considering a distribution~$P(g_m)$ for the
varying coupling strength of the~$m^{\rm th}$ 2LA. For the
multi--atom system, each atom experiences a varying coupling strength.
We assume a cut--off of~$N$ atoms in the system; more 
than~$N$ atoms introduces a negligible effect to measurable quantities.
We can express the coupling strength for the~$N$ atoms
by the multivariable coupling--strength distribution~$P(\vec{g})$
for~$\vec{g}\equiv(g_1,g_2,\dots, g_N)$. The Hamiltonian for~$N$ 2LAs
in the cavity is given by
\begin{equation} 
\label{H} 
H_N(\vec{g})= H^{(0)}+\sum_{m=1}^{N} H^{(m)}(g_m).
\end{equation}
Under the unitary 
transformation
\begin{equation}
\label{eq:unitr}
U_N(\vec{g})=U^{(N)}(g_N)\otimes U^{(N-1)}(g_{N-1}) \cdots
U^{(2)}(g_2)\otimes U^{(1)}(g_1),
\end{equation}
equation~(\ref{H}) reduces to 
\begin{equation}
\label{H:mult}
U^{\dag}_N(\vec{g})H_N(\vec{g})U_N(\vec{g})=\hbar\omega a^{\dag} a
+\sum_{m=1}^N \hbar\omega\sigma_z^{(m)}
+i\hbar g_m  \left( a^{\dag}\sigma_-^{(m)}-a\sigma_+^{(m)}\right),
\end{equation}
where we assume that~$g_m$ is always real and positive. The assumption
is valid because~$w^{(2)}(t)$ in equation~(\ref{eq:corr}) is invariant under
unitary transformations by~$U_N(\vec{g})$.

The case~$H_{M<N}(\vec{g})$, corresponding to fewer than~$N$ atoms 
in the cavity, is implicit by setting one or more
values of~$g_m$ to be zero. 
For $N=1$~the multiatom Hamiltonian (\ref{H}) 
reduces to the
JC Hamiltonian~\cite{Jay63} with `dressed states' 
\begin{equation}
\label{eq:JCstates:ground}
|0\rangle \equiv |0 \rangle_{\rm cav} \otimes |{\tt g} \rangle 
\end{equation}
and
\begin{equation}
\label{eq:JCstates:exited}
|n \rangle_{\pm} \equiv \frac{i}{\sqrt{2}}
\left( |n-1\rangle_{\rm cav} \otimes |{\tt e} \rangle
	\pm i |n\rangle_{\rm cav}\otimes
	|{\tt g}\rangle \right).
\end{equation}
Here 
$ \left \{ |{\tt g}\rangle ,\, |{\tt e}\rangle \right \}$ are
the ground and excited state of the 2LA. 
The~$N=2$ case is developed in detail in Appendix A. Instead of a ladder
consisting of a ground state as a singlet and higher--order states as
doublets, there is a singlet, then a triplet (for $n=1$) and then
quadruplets (for $n>1$). This assumes that the two atoms experience
different coupling strengths~$g_1\neq g_2$. For~$g_1=g_2$, the 
two--2LA Tavis--Cummings model is obtained~\cite{Tav69}.

\subsection{The master equation}
\label{subsec:master}

The Hamiltonian evolution (\ref{H:mult}) does not account for either
irreversible dynamics or contributions from the bichromatic driving 
field~${\cal E}(t)={\cal E}_1 \exp{(-i\omega_1 t)}
+{\cal E}_2 \exp{(-i\omega_2 t)}$
(with the resonance condition~$\omega_1=\omega-g_f$). To include all
these effects, we construct the quantum master equation.
The Born--Markov approximation is applied to both radiation reservoirs:
the reservoir for the field leaving the cavity and the reservoir
for direct fluorescence of the 2LA from the sides of the cavity.
The cavity damping rate is~$\kappa$,
and the emission rate into free space is~$\gamma_m$,
where~$\gamma_m$ is the inhibited spontaneous emission rate for the 
$m^{\rm th}$ atom. 
The master equation~\cite{San97} can be expressed as
$\dot{\rho} = {\cal L}\rho$ for~$\cal L$ the Liouvillean superoperator
and~$\rho$ the density matrix for the system.
More specifically the Liouvillean superoperator can be expressed 
as the sum of a time--independent Liouvillean operator,
a time--dependent Liouvillean operator and a `jump' term.
In the rotating 
frame, the time--independent non--Hermitian effective Hamiltonian is 
\begin{eqnarray}
\label{eq:Heff}
H_{\rm eff}({\vec{g}})=H_{\rm eff}^{(0)}+\sum_{m=1}^{N}  
H_{\rm eff}^{(m)}(g_m)
\end{eqnarray}	
with
\begin{eqnarray}
\label{Heff:terms}
H_{\rm eff}^{(m)}= \left\{ 
	\begin{array}{ll}
	\left( \omega - \omega_1 \right)a^{\dagger} a 
	- i\kappa a^{\dagger} a, \,\,\,\,\,\,\,
	\,\,\,\,\,\,\,\,\,\,\,\,\,\,\,\,
	\,\,\,\,\,\,\,\,\,\,\,\,\,\,\,\,\,\,\,\,\,\,\,\,\,\,
	\,\,\,
	 \mbox{if $m=0$} \\
	\\
	\left.\begin{array}{rr}
	\left( \omega - \omega_1 \right) \sigma^{(m)}_z 
	+ i g_m  
	(a^{\dagger} \sigma^{(m)}_- - a \sigma^{(m)}_+ ) \\	
	+\Upsilon^{(m)}({\cal E}_1)
	- i(\gamma_m/2) \sigma^{(m)}_+ \sigma^{(m)}_-
  	\end{array}\right\}\mbox{  $\;m>0$.}
	\end{array} \right.
\end{eqnarray}	
The monochromatic driving term is given by
\begin{equation}
\label{eq:Upsilon}
\Upsilon^{(m)}
({\cal E}_1)\equiv i {\cal E}_1 (\sigma^{(m)}_+ - \sigma^{(m)}_- ).
\end{equation}

Thus, the Liouville operator for the~$N$--atom cavity system is
\begin{equation}
\label{L}
{\cal L}\rho(\vec{g})=-\frac{i}{\hbar} \sum_{m=1}^{N}\bigg( 
H_{\rm eff}^{(m)} \rho(\vec{g})
-\rho(\vec{g}) H_{\rm eff}^{(m)\,\dag}\bigg)	
+{\cal D}(t;g)\rho+{\cal J}\rho
\end{equation}
with time--dependent driving term  
\begin{equation}
\label{L(t)}
{\cal D}(t;\delta) \rho
	= 
-i\left[ \Upsilon({\cal E}_2 e^{-i\delta t}) , \rho(\vec{g}) \right],
\end{equation}
and jump term 
\begin{equation}
\label{jump}
{\cal J} \rho = 2 \kappa a \rho(\vec{g}) a^{\dagger} 
	+\sum_{m=1}^N\gamma_m \sigma^{(m)}_- \rho(\vec{g}) \sigma^{(m)}_+.
\end{equation}
We assume that~$\gamma\equiv\gamma_m$ is
identical for all atoms.
For atom number fluctuations, the 2PCR is the sum of 2PCRs for
each case: no atom, one atom, two atoms, and so on. We calculate the
2PCR for each case and add each 2PCR component 
to obtain the resultant~$w^{(2)}$,
the 2PCR of equation~(\ref{eq:twocorr}), up to an unimportant
scaling coefficient. However, the contribution to the 2PCR must be
weighted according to the probabilities~$\{ p_m \}$, with~$p_m$
the probability for~$m$ atoms being in the cavity. Thus, the total 2PCR is 
$w^{(2)}=\sum_{m=0}^{\infty} p_m w_m^{(2)}$.

\subsection{The density matrix}
\label{subsec:densitym}

The master equation is used to solve the density matrix~$\rho$ in the
long--time limit, well after transients have disappeared. However, 
the bichromatic driving field results in an oscillatory solution 
for~$\rho$; a time--independent stationary solution is not obtained
for this case. The long--time limit for~$\rho$ can be solved by
employing a Bloch expansion in terms of the bichromatic 
driving field frequency 
difference~$\delta\equiv\omega_2-\omega_1$,
\begin{equation}
\label{Bloch}
\rho(t,\vec{g}) = \sum_{k=0}^{\infty} 
\rho_k(t,\vec{g}) e^{-i k \delta t},
\end{equation}
with~$\rho_k(t,\vec{g})$ time--dependent matrices.
In the long--time limit, $\dot{\rho}_k \approx 0$, and
$\rho_k(t,\vec{g}) \rightarrow \rho_k(\vec{g})$ is time--independent.
As the photocount integration time is expected to be long compared to the 
frequency~$\delta$, it is reasonable to approximate $\rho(t,\vec{g})$
by truncating expansion~(\ref{Bloch}). For photocount 
integration times larger than~$\delta^{-1}$, we truncate~(\ref{Bloch})
at~$k=0$, yielding~$\rho(t,\vec{g})\approx\rho_0(\vec{g})$, which is
time--independent. Approximating the time--dependent density matrix by
the first (time--independent) term in the Bloch expansion is
consistent with single--atom 2PCS studied in reference~\cite{San97}.

\begin{figure}
\begin{picture}(200,280)(45,10)
\font\gnuplot=cmr10 at 10pt
\gnuplot
\rotatebox{90}{\resizebox{400pt}{470pt}{\includegraphics{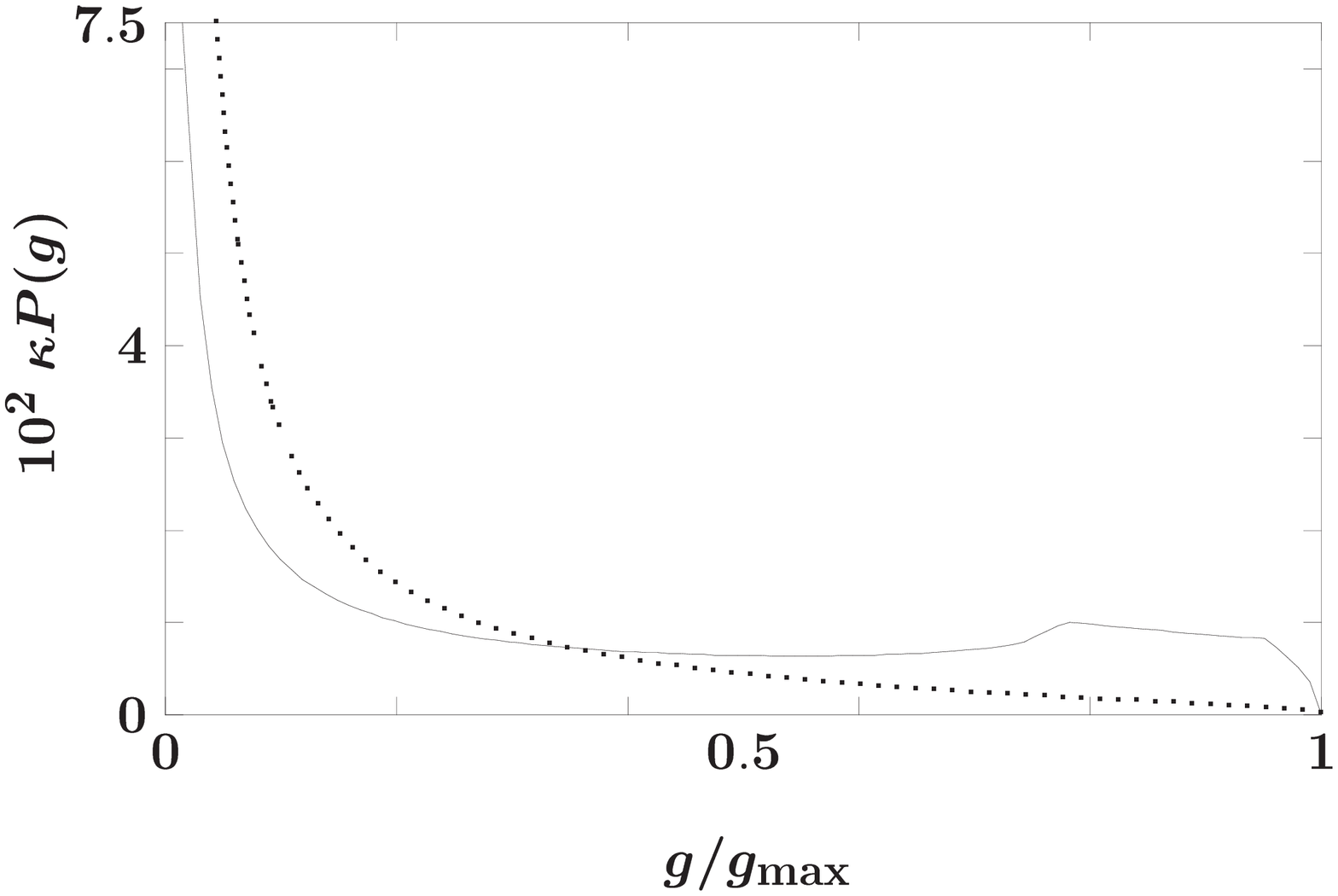}}}
\end{picture}
\caption{The scaled coupling strength distributions~$\kappa P(g)$ 
	as a function of~$g/g_{\rm max}$ for single atoms passing
	through an optical cavity supporting a single TEM$_{00}$ mode.
	The solid curve corresponds to a typical distribution for a
	rectangular mask filtering the atomic beam. The dotted line
	corresponds to the absence of a mask.}
\label{fig:P(g)}
\end{figure}

The resultant density matrix for the~$N$--atom cavity system is given by
\begin{equation}
\label{density:matrix}
\bar{\rho}(t)\approx
\bar{\rho}_0 \equiv \int_{Fg_{\rm max}}^{g_{\rm max}} P(\vec{g})
\rho_0(\vec{g}) 
d\vec{g}, 
\end{equation}
where $g_{\rm max}$ is the coupling strength at a cavity antinode,  
$Fg_{\rm max}$ 
is the effective lower bound cut--off for the coupling $(0<F<1)$
and~$P(\vec{g})=P(g_1)P(g_2)\cdots P(g_N)$ 
is the~$N$--atom coupling strength distribution, with two
typical plots of~$P(g)$ for one atom depicted in figure~\ref{fig:P(g)}
for~$\kappa$ the cavity loss rate~\cite{Car96,San97}.
The effect of averaging over~$P(\vec{g})$ is an inhomogeneous  
spectral broadening, and we use the overbar~$^{-}$ notation to 
denote averaging over~$P(\vec{g})$ to account for inhomogeneous
broadening. 
This broadening is due to atomic position variability.  
One distribution (dotted line) corresponds to the case of 
a uniformly distributed atomic beam entering the cavity and the other 
(solid line) to 
an atomic beam initially passing through a rectangular 
mask~\cite{San97}.
In both cases we have assumed a single--mode cavity
supporting a TEM$_{00}$ mode and atomic motion perpendicular to the
longitudinal axis of the cavity.

\section{Photon--coincidence spectroscopy for two atoms in the cavity}
\label{sec:twoatomspec}

Although PCS has been developed to probe the nonlinear portion of the
(one--atom) JC spectrum (2PCS probes the second couplet and multiphoton
coincidence spectroscopy~\cite{Hor99} probes higher levels), in this
section we show that 2PCS can also probe the nonlinear portion of
the spectrum for two--atom events. Interpreting the 2PCR for the
two--atom case is somewhat more complicated than for the JC spectrum
due to the presence of inhomogeneous
broadening for both atoms; methods for interpreting the 2PCR
peaks must therefore be more sophisticated. In this section, we describe
the method for performing PCS to extract a quantum field signature 
for two atoms in the cavity. 

\subsection{Excitation of two atoms coupled to a single cavity mode}
\label{subsec:twoats}

Two--atom dressed states are discussed in detail in Appendix~A,
and this description helps to understand the efficacy of PCS for
two--atom CQED. The strong--coupling regime implies that considering 
atoms and the cavity mode separately is not helpful: dressed states
are the preferred description.

The linear portion of the Hamiltonian spectrum consists of
the ground state and the
one--quantum triplet. This triplet reduces to a doublet for the
case that both atoms experience the same coupling 
strength~$g_1=g_2$~\cite{Tav69} and is the system being
probed in some  normal--mode (or vacuum Rabi) splitting
experiments which involve multiple atoms in the 
cavity~\cite{Rai89,Zhu90,Tho92}. 
Such experiments have been described well 
by semiclassical theories~\cite{Car94}.
The purpose of 2PCS is to extend beyond the linear regime to where
quantum field theory is essential for describing observations.
Thus, for the two--atom case, 2PCS should be designed to probe the
first quadruplet of states (or higher, as an analogy to multiphoton
coincidence spectroscopy~\cite{Hor99}).

The singlet, triplet and first quadruplet of dressed states for the
system consisting of two 2LAs coupled to a single cavity mode
are depicted in figure~\ref{fig:ladder}(b). Two--photon excitation to
the first quadruplet is developed by analogy with two--photon
excitation of the JC system to the second doublet, depicted
in figure~\ref{fig:ladder}(a), and discussed in references~\cite{Car96,San97}. 

For the (single--atom) JC system in figure~\ref{fig:ladder}(a), we depict the 
	two--photon excitation scheme from the ground state 
	$\left\vert 0 \right\rangle$ to the first two excited couplets
	$\left\vert n \right\rangle_{\varepsilon}$ 
	($n\in\{1,2\}$, $\varepsilon \in \{-,+\}$) of the dressed states.
	This single--atom system is a useful guide to understanding the
	two--atom scheme, and we briefly explain the single--atom case here.
	The challenge is to overcome inhomogeneous broadening of the 
	couplets~$\left\vert 1 \right\rangle_\varepsilon$ 
	and~$\left\vert 2 \right\rangle_\varepsilon$ 
	is~$2\hbar g_{\rm max}$ and~$2\sqrt2 \hbar g_{\rm max}$, 
	respectively, due largely to fluctuations in the atomic position. 
	Two two--photon excitations to the second couplet 
	of the (one--atom) JC system are
	depicted for a bichromatic driving field with one component
	of amplitude~${\cal E}_1$ and frequency~$\omega_1$ 
	and the other with amplitude~${\cal E}_2$
	and frequency~$\omega_2$.
	The excitation pathway on the right of figure~\ref{fig:ladder}(a)
	($\omega_1$ then~$\omega_2$)
	excites resonantly from~$\vert 0\rangle$ to~$\vert 1\rangle_-$
	and then may excite resonantly from~$\vert 1\rangle_-$ to either
	of the states~$\vert 2\rangle_{\pm}$.
	The excitation pathway on the left ($\omega_2$ then~$\omega_1$)
	excites resonantly from~$\vert 0\rangle$ to~$\vert 1\rangle_+$
	to~$\vert 2\rangle_-$ for~$g=(\sqrt2-1)g_f$.

	For 2PCS, the signature of entanglement is obscured 
	by the background 2PCR due to two~$\omega_2$ photons 
	contributing to excitation to the second couplet via 
	nonnegligible, off--resonant transitions. The method
	for overcoming this problem is called `background subtraction'.
	In this technique, the experiment is repeated twice, once
	with the bichromatic driving field and again with the
	fixed field turned off. The difference between these two 2PCRs
	shows all the features of the desired 2PCR without the deleterious
	effects of two~$\omega_2$ transitions to the second couplet 
	states~\cite{Car96,San97}.  

	Now we consider the case of two 2LAs atoms coupled to
	a single quantised field mode, 
	which is depicted in figure~\ref{fig:ladder}(b).
	Two--photon excitation occurs from the ground state 
	$\left\vert 0 \right\rangle$ via the triplet
	$\left\vert 1 \right\rangle_{\eta}$ to the first
	quadruplet
	 	$\left\vert 2 \right\rangle_{\varepsilon\,\varepsilon'}$
	(with $\eta\in\{-,0,+\}$ and 
	$\{\varepsilon,\varepsilon'\} \in \{-,+\}$) 
	with~$g_1\neq g_2$ assumed (the case~$g_1=g_2$ is 
	sufficiently unlikely to be ignored).
	Two two--photon excitation pathways to the second couplet are
	depicted for a bichromatic driving field with one component
	of amplitude~${\cal E}_1$ and the other with amplitude~${\cal E}_2$.
	The excitation pathway on the right ($\omega_1$ then~$\omega_2$)
	excites resonantly from~$\vert 0\rangle$ to~$\vert 1\rangle_-$
	and hence may excite resonantly to any 
	state of the
	set~$\{\vert 2\rangle_{\varepsilon \,\varepsilon'}\}$.
	The excitation pathway on the left ($\omega_2$ then~$\omega_1$)
	excites resonantly from~$\vert 0\rangle$ to~$\vert 1\rangle_{\eta}$
	to~$\vert 2\rangle_{\varepsilon \,\varepsilon'}$. By analogy
	with one--atom 2PCR, we can employ a spectral hole--burning
	approach for excitation to a subensemble of the first
	quadruplet and thereby operate in the nonlinear regime.

\subsection{Two--photon coincidence spectrum}
\label{subsec:twophcons}

For 2PCR of the JC system, the bichromatic driving field is used to
excite selectively to subensembles of the inhomogeneous broadened
first couplet of the JC ladder, as depicted in figure~\ref{fig:ladder}(a).
The second photon excites to a state of the second couplet only when the
resonance condition for~$\omega_1+\omega_2$ equals~$2\omega\pm\sqrt{2}g$
for the selected subensemble with coupling strength~$g$; this 
resonance signature provides unambiguous evidence of a quantum field
effect through the~``$\sqrt2$''.

For 2PCR of the two--atom plus cavity mode system, selecting a particular
subensemble corresponding to a fixed~$\vec{g}=(g_1,g_2)$ 
is not possible. To see how
this complication arises, consider the case that~$\omega_1$ is tuned
such that direct excitation occurs from~$\vert 0\rangle$ 
to~$\vert 1\rangle_-$, which is given by equation~(\ref{eq:1_pm}), for 
certain choices of~$g_1$ and~$g_2$. The energy of  
state~$\vert 1\rangle_-$ corresponds to the eigenvalue~$\lambda_-$ of
equation~(\ref{eq:pm}). Choosing a particular~$\omega_1$
fixes~$\tilde{g}\equiv\sqrt{g_1^2+g_2^2}$ according
to the expression~$\lambda^{-}=\omega-\tilde{g}$
but does not fix~$g_1$ and~$g_2$ separately. Therefore, choosing a 
particular~$\omega_1$ does not select a unique subensemble but
rather a class of subensembles corresponding to the 
constraint~$\tilde{g}=\omega-\lambda^{-}=\omega-\omega_1$.

This large class of subensembles is not too difficult to manage, however. 
We are particularly interested in factorisable distributions of the
type~$P(\vec{g})=P(g_1, g_2)=P(g_1)P(g_2)$; 
i.e.\ $P(g_m)$ is identical for each
independent atom, and~$P(g_i)$
appears as shown in figure~\ref{fig:P(g)}. Choosing a fixed~$\omega_1$
constrains the selected subensembles to values of~$\tilde{g}$ in the
domain of interest, namely large~$\tilde{g}$, and effectively reduces 
contributions such as high~$g_1$ and very low~$g_2$ that would negate
the desired 2PCR peaks. 

\begin{figure}
\begin{picture}(200,430)(20,-240)
\font\gnuplot=cmr10 at 10pt
\gnuplot
\rotatebox{90}{\resizebox{400pt}{420pt}{\includegraphics{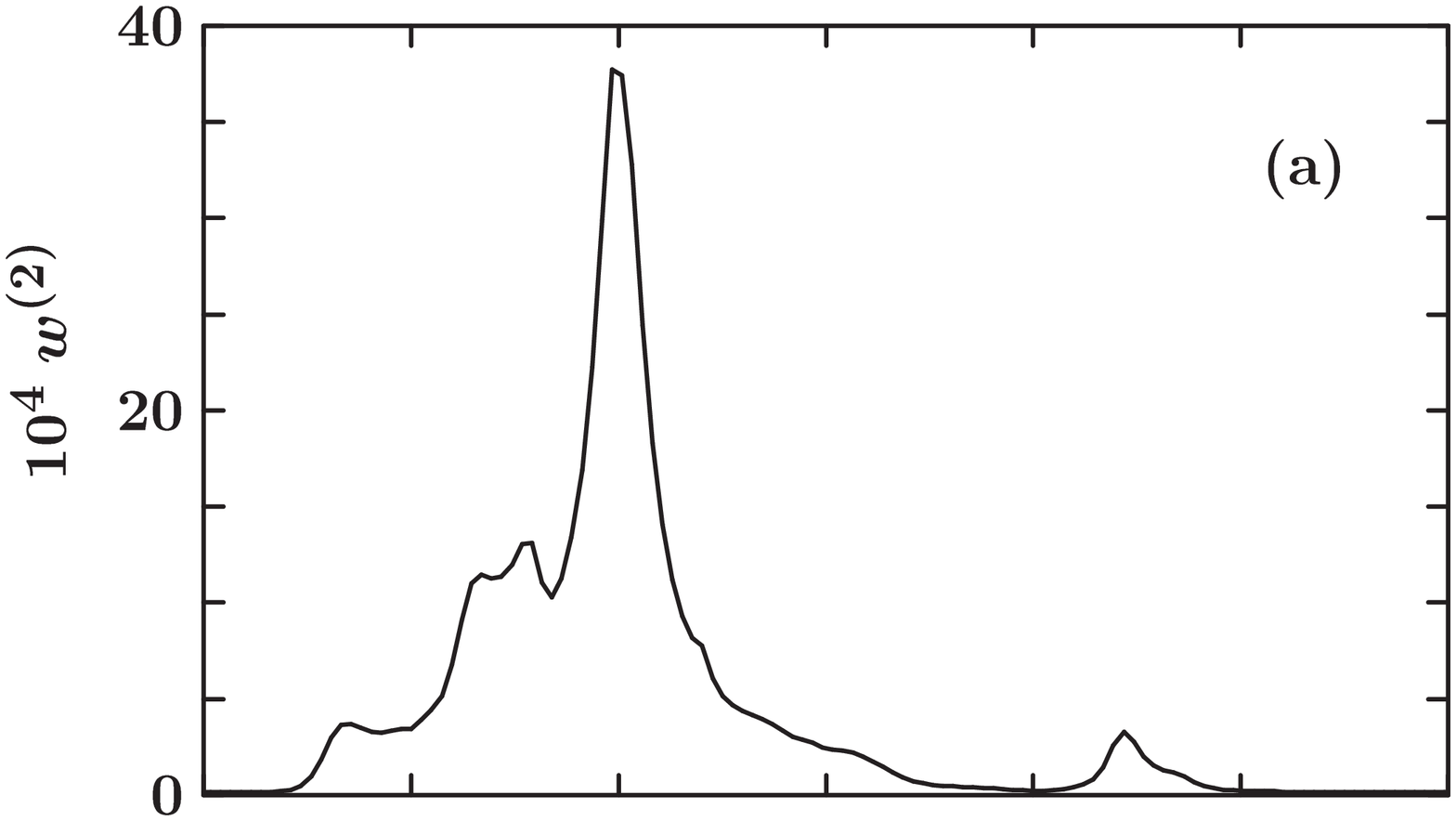}}}
\end{picture}
\begin{picture}(200,330)(220,10)
\font\gnuplot=cmr10 at 10pt
\gnuplot
\rotatebox{90}{\resizebox{400pt}{420pt}{\includegraphics{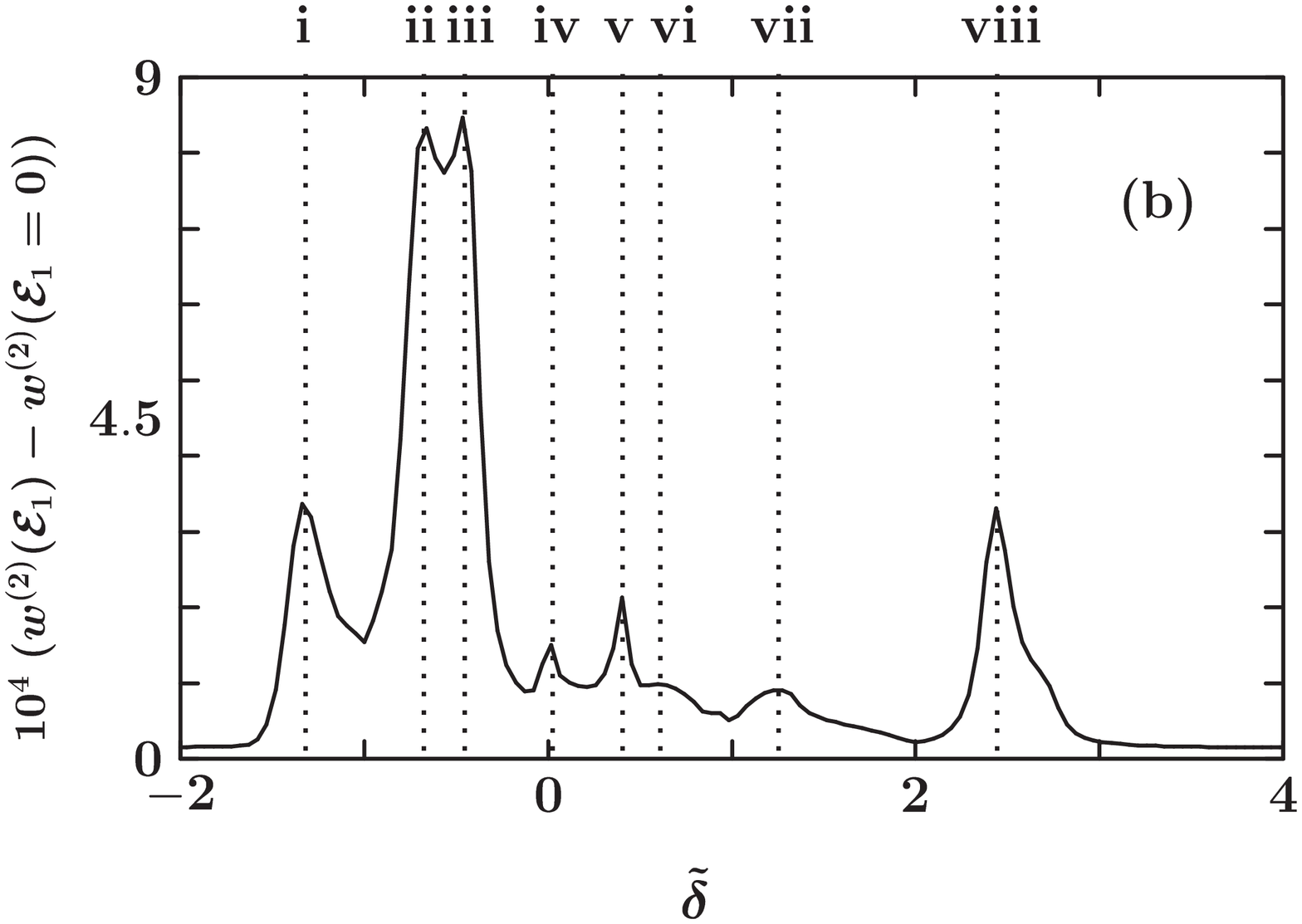}}}
\end{picture}
\caption{Two--photon count rate vs normalised 
	scanning 
	frequency for  two atoms in 
	a sparse atomic beam for
	~$\tilde{g}=g_f=63\kappa$, ${\cal E}_1/\kappa=1/\sqrt2$,
	${\cal E}_2/\kappa=\sqrt{2}$ and $\gamma/\kappa=2$
	(a)~without and (b)~with background subtraction. 
	Vertical dotted lines are placed in (b)~to 
	identify the obvious peaks. 
	}
\label{fig:smeared_gf63} 
\end{figure}

For 2PCR of the two--atom--plus--cavity system, the 2PCR is enhanced, not
for specific~$g_1$ and~$g_2$ but rather for a range of~$g_1$ and~$g_2$
constrained only by~$\tilde{g}$ being determined by the choice of~$\omega_1$
(and we unfortunately cannot fix both~$g_1$ and~$g_2$ in experimental
circumstances).
This means that the expected 2PCR peaks are broadened by this limited
control of~$g_1$ and~$g_2$. The 2PCR for two atoms in the cavity, 
with strong coupling to the mode and a weak bichromatic driving field,
is plotted in figure~\ref{fig:smeared_gf63}.

As an example, 
we consider the outlying strong peak at~$\tilde\delta\doteq 2.414$
(peak~$viii$ of figure~\ref{fig:smeared_gf63}(b)). As
we will show, this peak is due to the excitation 
pathway~$\vert 0\rangle\longleftrightarrow\vert 1\rangle_-
\longleftrightarrow \vert 2\rangle_{++}$. The frequency difference
for~$\vert 0\rangle\longleftrightarrow\vert 1\rangle_-$ is~$\lambda^{-}$,
as discussed above, and the transition for~$\vert 1\rangle_-
\longleftrightarrow \vert 2\rangle_{++}$ is~$\lambda_1^{++}-\lambda^{-}$. The
expression for normalised detuning ($g_f\equiv \omega-\omega_1$),
\begin{equation}
\label{eq:deltatil}
\tilde\delta=\frac{\omega_2-\omega}{\omega-\omega_1}=
g_f^{-1}(\lambda_1^{++}-\lambda^{-}-\omega),
\end{equation}
could be plotted as a line in the three--dimensional space
spanned by the~$g_1$,~$g_2$ and~$\tilde\delta$ axes. We do not
present such a plot here but rather note that the trajectory intersects
the observed peak value~$\tilde\delta\doteq 2.414$ (peak {\em viii}
of figure~\ref{fig:smeared_gf63})
for~$g_1=g_{\rm max}$ and~$g_2=0$ or vice versa. This intersection tells us
that the peak at~$\tilde\delta\doteq 2.414$ is really just a 
single--atom JC system peak because the second atom is effectively
decoupled (negligible~$g_2$). 
Setting the frequency~$\omega_1$ to~$\lambda^{\pm}$ is therefore
not the best way to probe two--atom effects because~$\lambda^{\pm}$
depends on~$\tilde{g}$ but is not dependent on~$g_1$ and~$g_2$
separately. This has the effect of producing a peak for the
case that~$\vert 1\rangle_{\pm}$ reduces (for~$g_1=\tilde{g}$,~$g_2=0$) 
to~$2^{-1/2}(\vert 0\rangle\vert {\tt e}\rangle
\pm\vert 1\rangle\vert {\tt g}\rangle)\vert {\tt g}\rangle$. Atom 2
remains in the ground state. Similarly, setting~$\omega_1$ 
to~$\lambda^0$ has the problem that it is completely independent 
of~$\omega_1$ and~$\omega_2$.

\begin{figure}
\begin{picture}(200,280)(150,250)
\font\gnuplot=cmr10 at 10pt
\gnuplot
\rotatebox{0}{\resizebox{650pt}{680pt}{\includegraphics{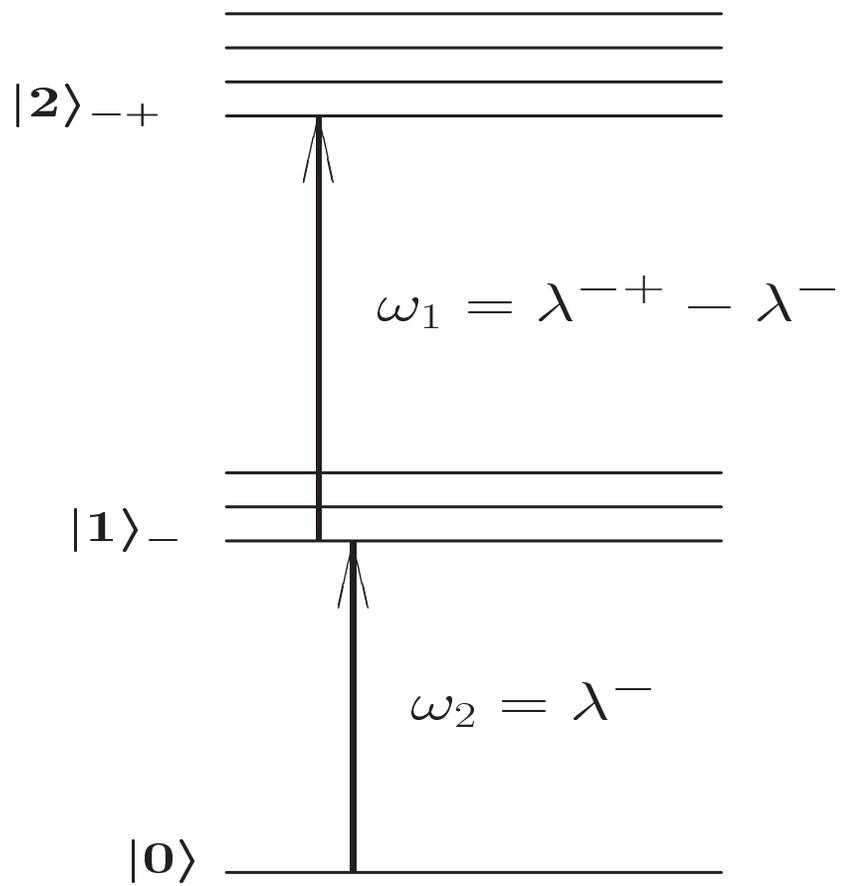}}}
\end{picture}
\caption{The excitation pathway responsible for the two--atom signature in
	2PCS.  
	}
\label{fig:omega2->1} 
\end{figure}

The dilemma of not having a good choice for~$\omega_1$ to
excite from~$\vert 0\rangle$ to~$\vert 1\rangle_{\pm}$ or~$\vert 1\rangle_0$
is solved by setting~$\omega_2$ to~$\lambda^{\pm}$ or~$\lambda^{0}$.
However,~$\omega_2$ is the scanning frequency. The proposal here, though,
is first to assume that~$\omega_2$ is set to the frequency~$\lambda^{-}$,
which excites~$\vert 0\rangle\longleftrightarrow\vert 1\rangle_-$.
Then~$\omega_1=\lambda^{-+}-\lambda^{-}$, 
which is appropriate for the 
transition~$\vert 1\rangle_-\longleftrightarrow\vert 2\rangle_{-+}$.
This counter--intuitive choice (letting the scanning frequency excite
the first transition rather than the second) enables peak~$i$ of
figure~\ref{fig:smeared_gf63} to be understood.
The 2PCR would then be enhanced for~$\omega_2=\lambda^{-}$ with $\omega_1$
fixed. The peak corresponding to this excitation pathway 
(depicted in figure~\ref{fig:omega2->1}) is 
observed at~$\tilde\delta=-1.345$ (peak {\em i} of 
figure~\ref{fig:smeared_gf63}). Substituting this
value of~$\tilde\delta$ into the expression for~$\tilde\delta(g_1,g_2)$
obtained in Appendix~A yields the two solutions~$g_1/\kappa=64.9$ 
and~$g_2/\kappa=45.5$
and the reverse. The peak at~$\tilde\delta=-1.345$ has its most
significant contribution from the case that both atoms contribute
to the 2PCR signal.

This analysis helps to understand the 2PCR peak structure depicted in
figure~\ref{fig:smeared_gf63}, but a detailed analysis is afforded
by the method of suppressed transitions developed in reference~\cite{Hor00a}.
In this method we identify the specific transitions contributing to
each peak.

\subsection{Method of suppressed transitions}
\label{subsec:msuptrans}

We apply the method of suppressed transitions to understand 
figure~\ref{fig:smeared_gf63} and to identify the peaks corresponding
to genuine quantum field effects for two atoms in the cavity.
In the method of suppressed transitions~\cite{Hor00a}, 
we modify the Liouvillean superoperator~${\cal L}$ in numerical 
simulations
by artificially
eliminating particular driving terms responsible for certain peaks. 
The isolation of specific transitions is obtained by the following
procedure. For example, let us consider the influence on the 2PCR
of the~$\vert 0\rangle\longleftrightarrow\vert 1\rangle_-$ transition.
We can write the effective Hamiltonian~(\ref{eq:Heff}) for two 2LAs
as a matrix in the dressed--state basis. To isolate the influence
of the~$\vert 0\rangle\longleftrightarrow\vert 1\rangle_-$
transition,
we can set the matrix element~$\langle 0\vert \Upsilon({\cal E}_1)
\vert 1\rangle_-$ and its complex
conjugate~$_-\langle 1\vert \Upsilon({\cal E}_1)
\vert 0\rangle$ to be zero where~$\Upsilon({\cal E}_1)$ 
is the (Hermitian)
driving term~$\Upsilon({\cal E}_1)=\sum_{m=1}^N \Upsilon^{(m)}({\cal E}_1)$
for~$\Upsilon^{(m)}({\cal E}_1)$ defined by~equation~(\ref{eq:Upsilon}).
In addition the matrix element~$\langle 0\vert \Upsilon({\cal E}_2 \exp{(-i\delta t)})\vert 1\rangle_-$
and its complex conjugate can both be set to zero. The jump
term~${\cal J}$ is not modified because only the driving terms and
their effects are of concern in this analysis of suppressed transitions.

In subsection~\ref{subsec:twophcons}, 
we have shown that transitions from~$\vert 0\rangle$ 
to~$\vert 1\rangle_\eta$ ($\eta \in \{-,0,+\}$) via an~$\omega_1$ photon
is not the best choice for probing the two--2LA cavity system as these
excitations lead to single--atom JC system peaks.
The artificial removal of these transitions 
by the method of suppressed transitions eliminates the
(single--atom) JC system peaks in the simulations and 
thereby helps to identify 
those peaks that are primarily due to two--atom cavity events.
The method of suppressed transitions is a mathematical tool for
interpreting peaks in the two--photon spectral structure, not
a physical process.

\begin{figure}
\begin{picture}(200,250)(80,110)
\font\gnuplot=cmr10 at 10pt
\gnuplot
\rotatebox{90}{\resizebox{370pt}{500pt}{\includegraphics{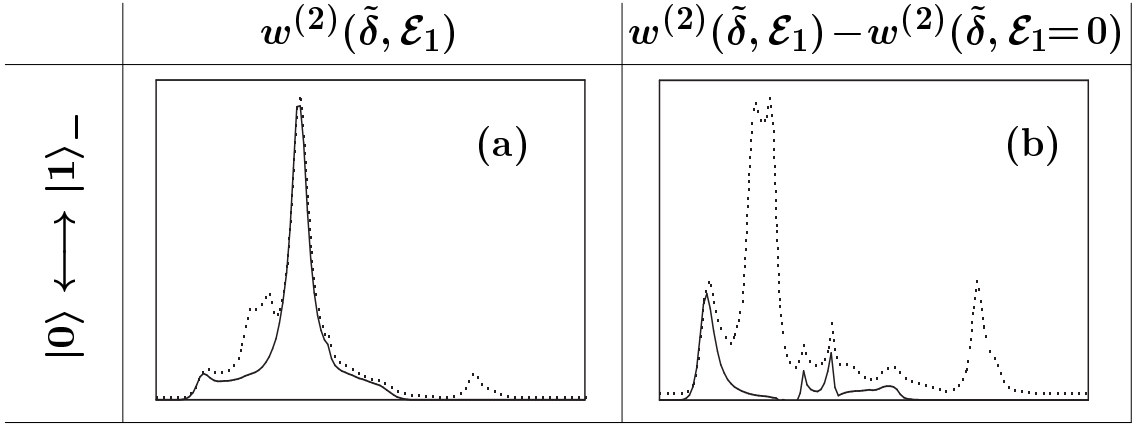}}}
\end{picture}
\caption{The 2PCR (a) without and (b) with background subtraction
	as a dotted line. The solid line is the 2PCR for 
	the~$\omega_1$--driven
	$\vert 0\rangle\longleftrightarrow\vert 1\rangle_-$
	transition  
	artificially suppressed in the simulations.
	The dotted lines are replicates of those in 
	figure~\ref{fig:smeared_gf63}(a,~b), and the scales here
	correspond to those scales.
	}
\label{fig:sparsebeam:ef01} 
\end{figure}

\begin{figure}
\begin{picture}(200,250)(-50,0)
\font\gnuplot=cmr10 at 10pt
\gnuplot
{\resizebox{600pt}{700pt}{\includegraphics{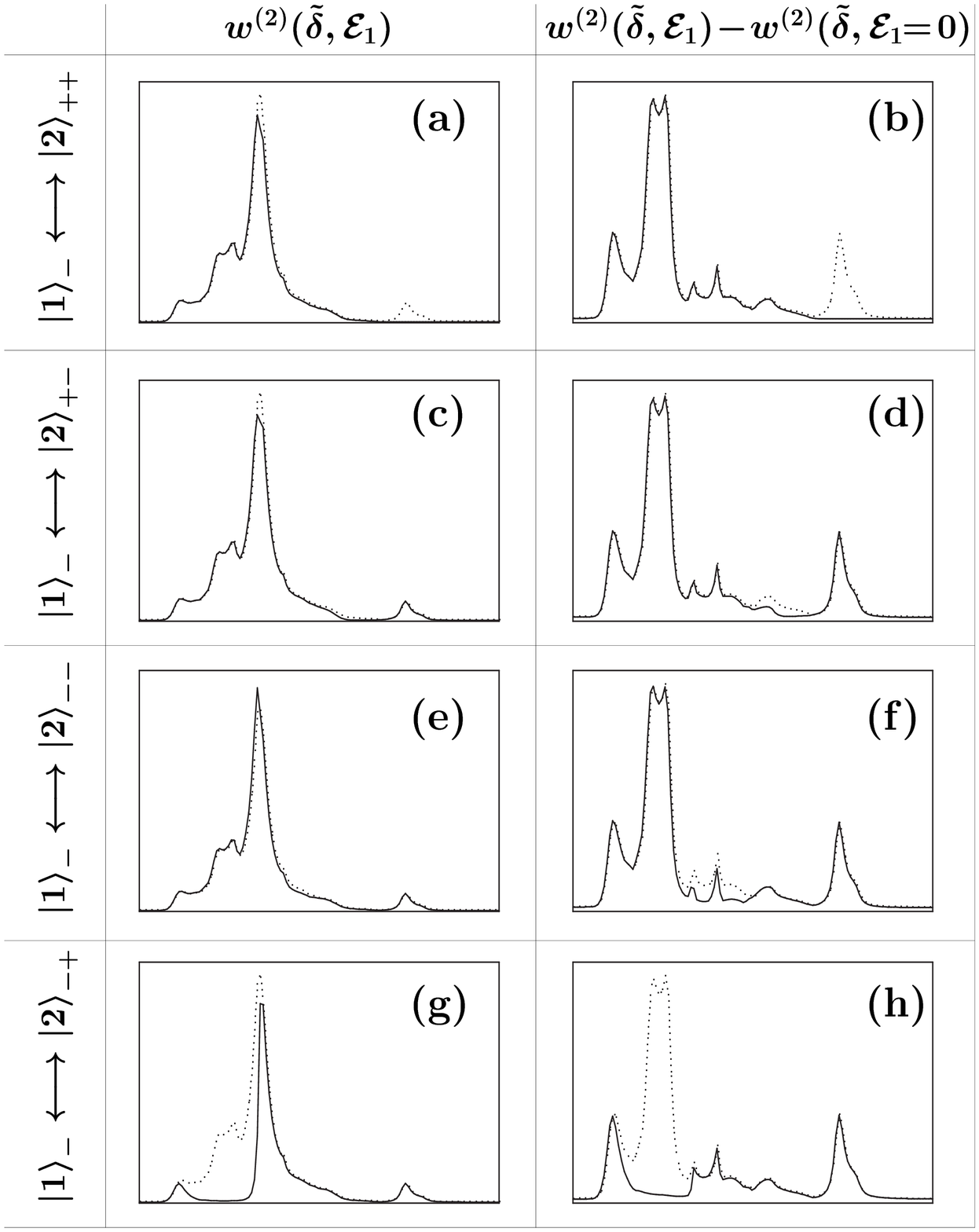}}}
\end{picture}
\caption{
	The 2PCR without (first column) and  with (second column) 
	background subtraction
	as the dotted line. The solid line 
	corresponds to the 2PCR for an~$\omega_2$--driven 
	transition artificially
	suppressed in the numerical simulation. The suppressed
	transition 
	is ~$\vert 1\rangle_-\longleftrightarrow\vert 2\rangle_{++}$
	for the first row,~$\vert 1\rangle_-\longleftrightarrow
	\vert 2\rangle_{+-}$ for the second row,~$\vert 1\rangle_-
	\longleftrightarrow\vert 2\rangle_{--}$ for the third row
	and~$\vert 1\rangle_-\longleftrightarrow
	\vert 2\rangle_{-+}$ for the fourth row.	
	}
\label{fig:sparsebeam:es12} 
\end{figure}

We begin by setting both 
the matrix element~$\langle 0\vert \Upsilon({\cal E}_1)
\vert 1\rangle_-$ and its complex conjugate to zero
as this transition is present in every single--atom peak, and
suppressing this transition is an excellent first step to identify
peaks associated with two atoms coupled to the cavity mode.
The result is depicted in figure~\ref{fig:sparsebeam:ef01}(a) 
without and~\ref{fig:sparsebeam:ef01}(b) with 
background subtraction. We observe that
eliminating these two matrix elements causes a dramatic reduction
of peaks~{\em ii}, {\em iii}, {\em vi}, {\em vii}, {\em viii} 
with the labelled peak structure of figure~\ref{fig:smeared_gf63}(b)
replicated in figure~\ref{fig:sparsebeam:ef01}(b) as a dotted line. 
The reason for these reductions is that
the excitation paths responsible for these peaks all have
a significant contribution from the
~$\vert 0\rangle\longleftrightarrow\vert 1\rangle_-$ transition, 
employing a photon of frequency~$\omega_1$.
Suppressing~$\langle 0\vert \Upsilon({\cal E}_1)\vert 1\rangle_-$ and
its conjugate only supplies a partial interpretation of the peak structure.
A further understanding of these peaks is obtained by imposing 
the condition
that~$_-\langle 1\vert \Upsilon({\cal E}_2\exp(i \delta t)) 
\vert 2\rangle_{++}$
and its complex conjugate are zero. As shown in 
in figure~\ref{fig:sparsebeam:es12},
photon~$\omega_2$ dominates 
the~$\vert 1\rangle_-\longleftrightarrow\vert 2\rangle_{++}$ transition
responsible for peak {\em viii}. 
Thus, the~$\omega_1$--driven excitation
pathway~$\vert 0\rangle\longleftrightarrow\vert 1\rangle_-$
followed by 
 an~$\omega_2$--driven transition to the~$\vert 2\rangle_{++}$ state
is responsible for peak~{\em viii} as shown in 
figures~\ref{fig:sparsebeam:ef01}(b) 
and~\ref{fig:sparsebeam:es12}(b), respectively. 
This method of 
suppressed transitions validates the analysis of peak~{\em viii} in
subsection~\ref{subsec:twophcons}, where the excitation 
pathway~$\vert 0\rangle\longleftrightarrow\vert 1\rangle_-
\longleftrightarrow\vert 2\rangle_{++}$ was suggested as
being responsible for the existence of this 2PCR peak.

As another example, we consider the 
transition~$\vert 1\rangle_-\longleftrightarrow\vert 2\rangle_{+-}$.
By eliminating this transition, peak {\em vii} (and vicinity)
 are reduced as shown in figure~\ref{fig:sparsebeam:es12}(d). 
Thus, the~$\vert 0\rangle\longleftrightarrow\vert 1\rangle_-
\longleftrightarrow\vert 2\rangle_{+-}$ pathway, induced by an~$\omega_1$ 
photon
followed by an~$\omega_2$ photon, is partially
responsible for the increase of the 2PCR at {\em vii} (and vicinity).
Figures~\ref{fig:sparsebeam:ef01} 
and~\ref{fig:sparsebeam:es12}(f) show clearly that 
an $\omega_1$--driven 
$\vert 0\rangle\longleftrightarrow\vert 1\rangle_-$ transition
and an~$\omega_2$--driven 
$\vert 1\rangle_-\longleftrightarrow
\vert 2\rangle_{--}$ transition are responsible for the background of peaks
{\em iv}, {\em v} and {\em vi} but not for the actual peaks themselves. 
Setting~$_-\langle 1\vert \Upsilon({\cal E}_2 \exp{(i\delta t)})
\vert 2\rangle_{-+}$ and its complex conjugate to zero  
eliminates 
peaks {\em ii} and {\em iii} as shown in figures~\ref{fig:sparsebeam:es12}(g, h).
Clearly, peaks {\em ii} and {\em iii} are primarily due to 
a resonant~$\vert 0\rangle\longleftrightarrow \vert 1\rangle_-$
transition via an~$\omega_1$ photon, followed by an excitation
via an~$\omega_2$ photon
to~$\vert 2\rangle_{-+}$ state. Peaks {\em ii} and {\em iii} are the 
dominant peaks because, apart from the fixed field with
frequency~$\omega_1$ 
driving~$\vert 0\rangle \longleftrightarrow \vert 1\rangle_-$, the
scanning field also drives~$\vert 0\rangle\longleftrightarrow\vert 1\rangle_-$
via an off--resonance transition;
thus, higher populations of the~$\vert 2\rangle_{-+}$ is expected.

\begin{figure}
\begin{picture}(200,250)(-50,0)
\font\gnuplot=cmr10 at 10pt
\gnuplot
{\resizebox{600pt}{700pt}{\includegraphics{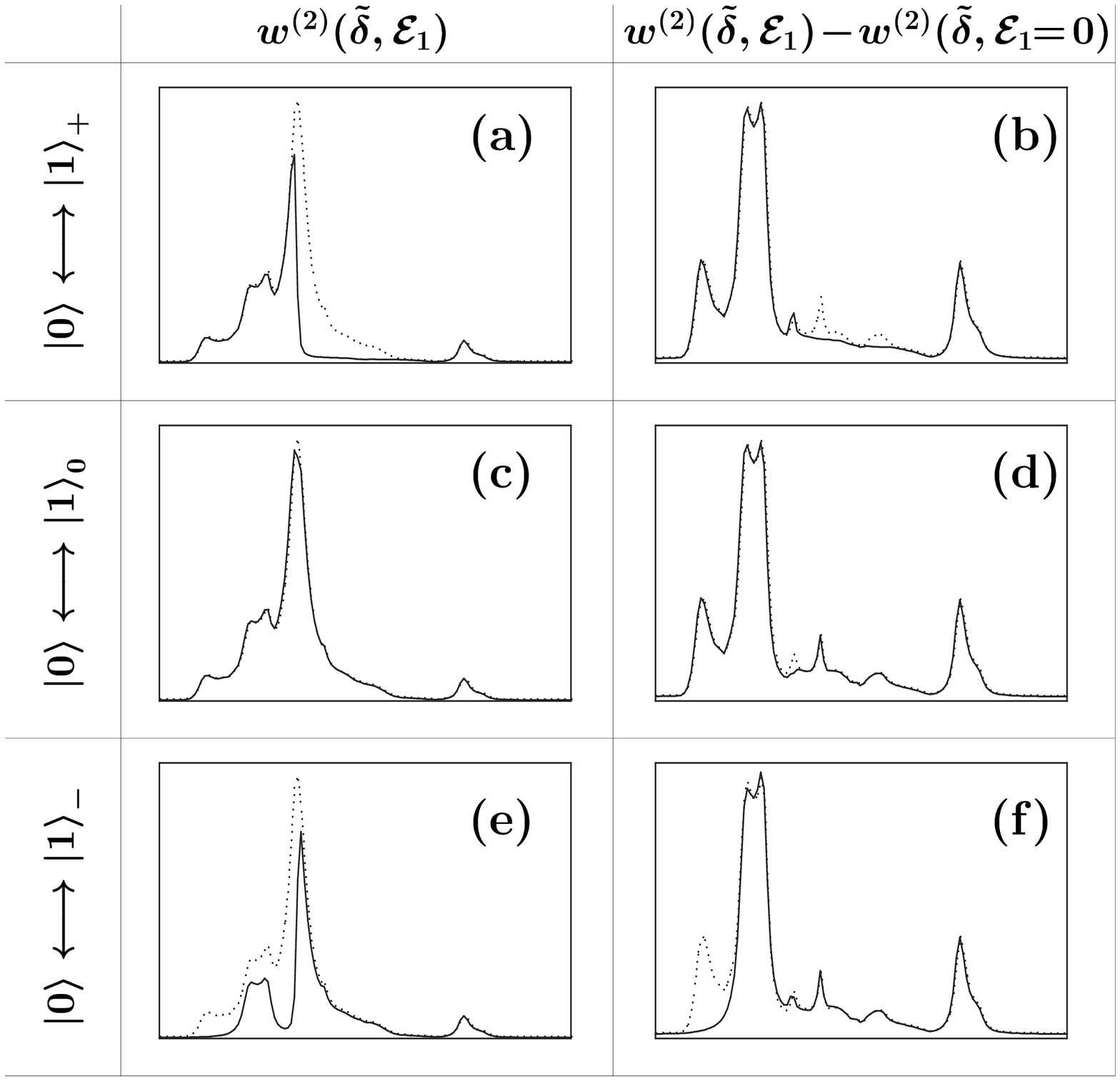}}}
\end{picture}
\caption{The 2PCR without (first column) and  with (second column) 
	background subtraction
	as the dotted line. The solid line 
	corresponds to the 2PCR for an~$\omega_2$--driven
	transition artificially
	suppressed in the numerical simulation. The suppressed
	transition 
	is ~$\vert 0\rangle\longleftrightarrow\vert 1\rangle_{+}$
	for the first row,~$\vert 0\rangle\longleftrightarrow
	\vert 1\rangle_{0}$ for the second row and~$\vert 0\rangle
	\longleftrightarrow\vert 1\rangle_{-}$ for the third row.
	}
\label{fig:sparsebeam:es01} 
\end{figure}

\begin{figure}
\begin{picture}(200,250)(-50,0)
\font\gnuplot=cmr10 at 10pt
\gnuplot
{\resizebox{600pt}{700pt}{\includegraphics{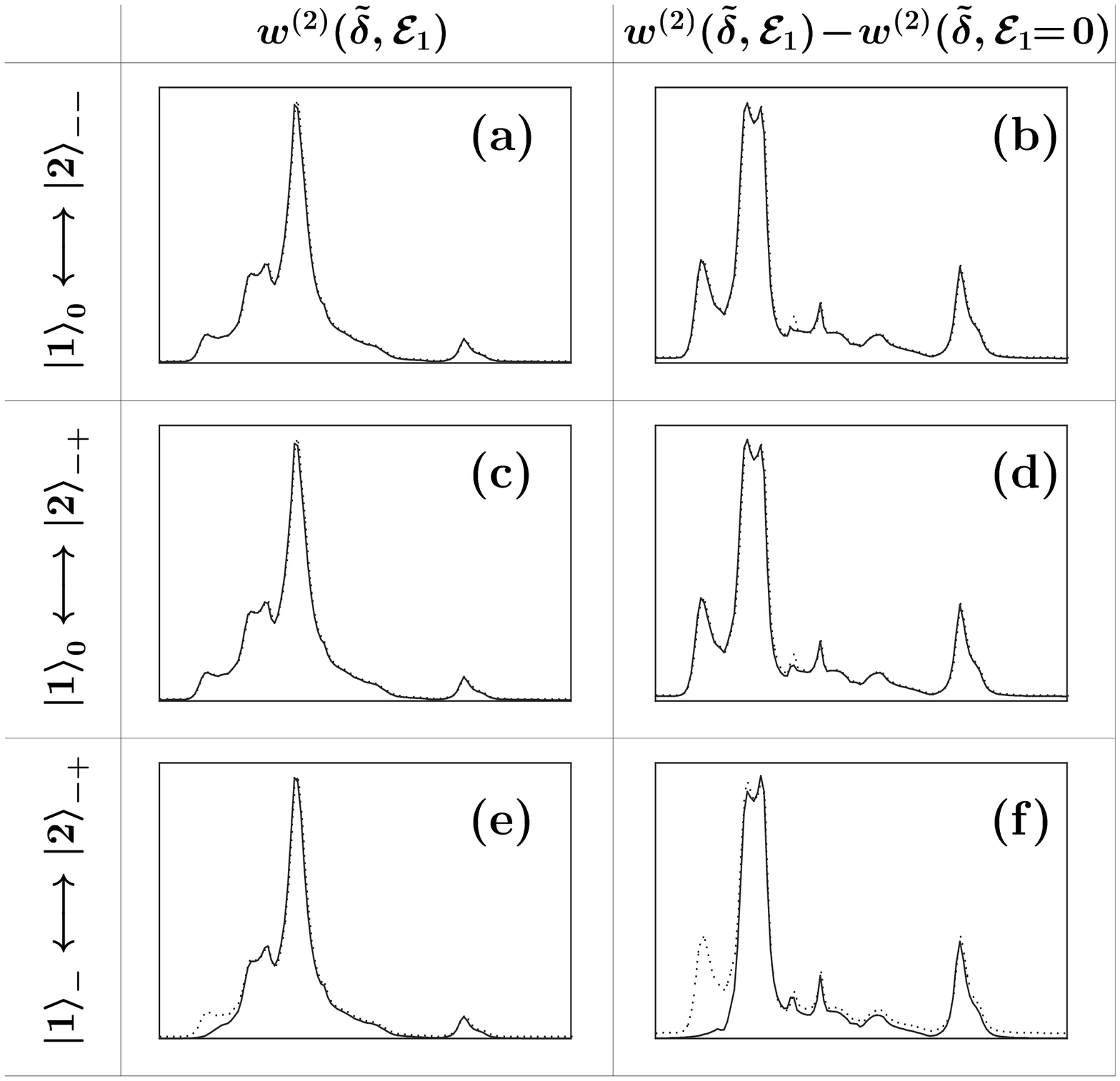}}}
\end{picture}
\caption{The 2PCR without (first column) and  with (second column) 
	background subtraction
	as the dotted line. The solid line 
	corresponds to the 2PCR for an~$\omega_1$--driven
	transition artificially
	suppressed in the numerical simulation. The suppressed
	transition 
	is ~$\vert 1\rangle_0\longleftrightarrow\vert 2\rangle_{--}$
	for the first row,~$\vert 1\rangle_0\longleftrightarrow
	\vert 2\rangle_{-+}$ for the second row and~$\vert 1\rangle_-
	\longleftrightarrow\vert 2\rangle_{-+}$ for the third row.
	}
\label{fig:sparsebeam:ef12} 
\end{figure}

\begin{figure}
\begin{picture}(200,250)(-50,0)
\font\gnuplot=cmr10 at 10pt
\gnuplot
{\resizebox{600pt}{700pt}{\includegraphics{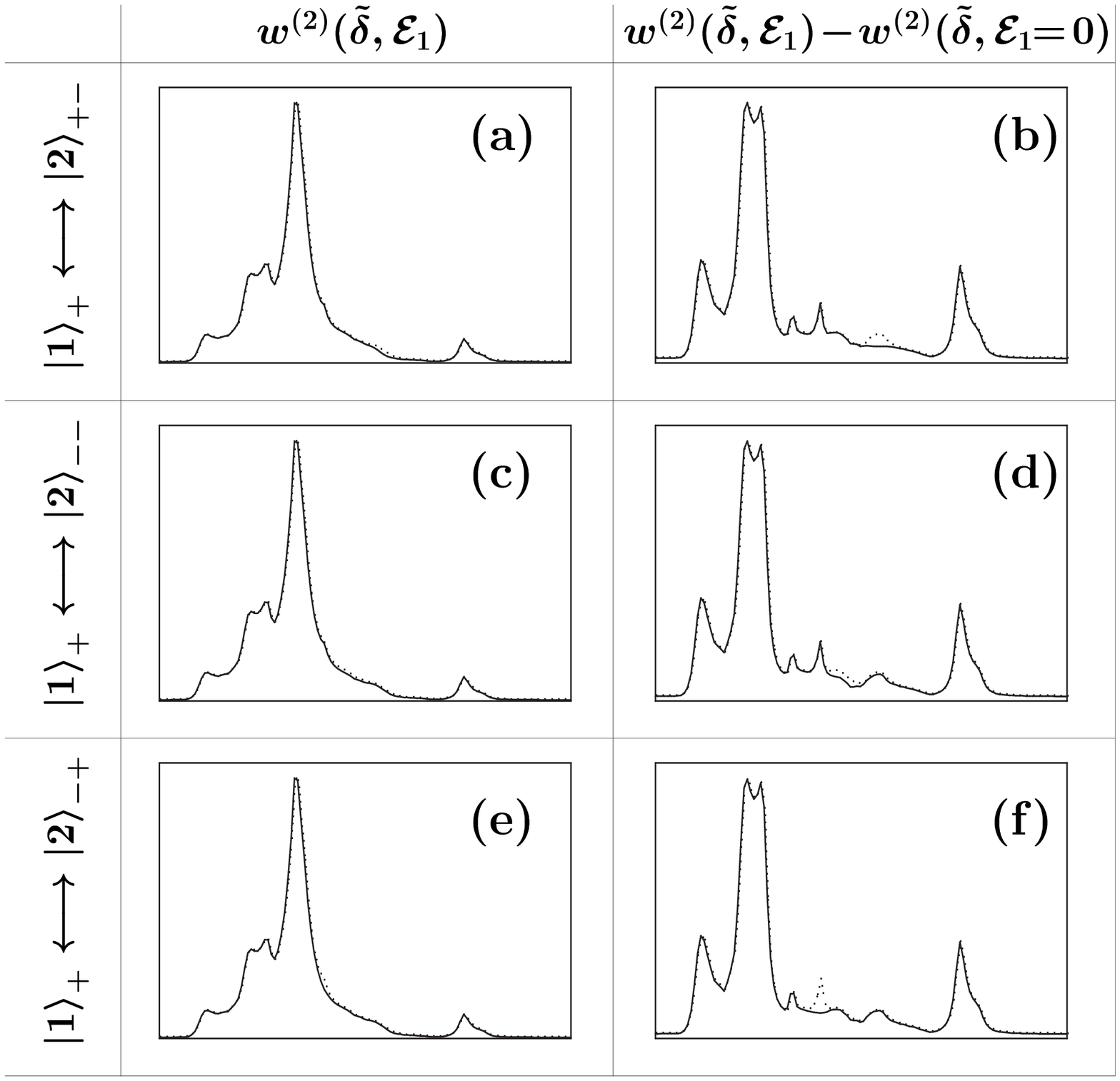}}}
\end{picture}
\caption{The 2PCR without (first column) and  with (second column) 
	background subtraction
	as the dotted line. The solid line 
	corresponds to the 2PCR for an~$\omega_1$--driven 
	a transition artificially
	suppressed in the numerical simulation. The suppressed
	transition 
	is ~$\vert 1\rangle_+\longleftrightarrow\vert 2\rangle_{+-}$
	for the first row,~$\vert 1\rangle_+\longleftrightarrow
	\vert 2\rangle_{--}$ for the second row and~$\vert 1\rangle_+
	\longleftrightarrow\vert 2\rangle_{-+}$ for the third row.
	}
\label{fig:sparsebeam:ef12_1} 
\end{figure}

Thus far, we have identified those peaks that are due
to single--atom events. The remaining peaks, particularly 
{\em iv}, {\em v} and {\em vii}, are the ones of interest because they
are primarily due
to the two--atom cavity signature, and this is because 
the transition~$\vert 0\rangle\longleftrightarrow\vert 1\rangle_\eta$ state
is via an~$\omega_2$ photon. 
To explain   
peaks {\em iv}, {\em v} and {\em vii}, we set
$\langle 0\vert \Upsilon({\cal E}_2 \exp(-i\delta t))\vert 1\rangle_{\eta}$ and
$_{\eta}\langle 1\vert \Upsilon({\cal E}_1)
\vert 2\rangle_{\varepsilon\,\varepsilon'}$,
and their complex conjugates, all  
to zero. For example, setting $\langle 0\vert \Upsilon({\cal E}_2 
\exp(-i\delta t))\vert 1\rangle_+$
and its complex conjugate to zero eliminates the peaks at {\em v}, {\em vi} 
and
{\em vii} as shown in figure~\ref{fig:sparsebeam:es01}(b).
With this result and figure~\ref{fig:sparsebeam:ef12_1}(b),
we observe 
that the peak at {\em vii} is due to an~$\omega_2$--driven 
$\vert 0\rangle\longleftrightarrow\vert 1\rangle_+$ transition
followed by an~$\omega_1$--driven 
$\vert 1\rangle_+\longleftrightarrow\vert 2\rangle_{+-}$ transition. 
Fixing $_+\langle 1\vert \Upsilon({\cal E}_1)
\vert 2\rangle_{--}$ and its complex conjugate to zero 
reduces the peak at~{\em vi} as shown in
figure~\ref{fig:sparsebeam:ef12_1}(d).
Thus, the small bump, namely peak {\em vi} 
(to the right of peak {\em v}), is due to 
an~$\omega_2$--driven 
excitation~$\vert 0\rangle\longleftrightarrow\vert 1\rangle_+$ 
followed by an~$\omega_1$--driven~$\vert 1\rangle_+
\longleftrightarrow\vert 2 \rangle_{--}$ transition.
As shown in  figure~\ref{fig:sparsebeam:es01}(b)
and figure~\ref{fig:sparsebeam:ef12_1}(f), 
the removal of the~$\omega_2$--driven
transition~$\vert 0\rangle\longleftrightarrow\vert 1\rangle_{+}$
and the~$\omega_1$--driven 
$\vert 1\rangle_+\longleftrightarrow\vert 2\rangle_{-+}$
transition yields a diminished peak {\em v}.
Peak {\em iv} is diminished if the transitions 
$\langle 0\vert \Upsilon({\cal E}_2 \exp(-i\delta t))\vert 1\rangle_0$,
$_0\langle 1\vert \Upsilon({\cal E}_2 
\exp(-i\delta t))\vert 2\rangle_{--}$ and
$_0\langle 1\vert \Upsilon({\cal E}_2 
\exp(-i\delta t))\vert 2\rangle_{-+}$, together with their complex
conjugates, are suppressed by being set to zero
as shown in figures~\ref{fig:sparsebeam:es01}(d),
\ref{fig:sparsebeam:ef12}(b) and~\ref{fig:sparsebeam:ef12}(d). 
Clearly, peak {\em iv} is due
to an~$\omega_2$--driven 
resonant pathway~$\vert 0\rangle\longleftrightarrow\vert 1\rangle_0$, 
followed by~$\omega_1$--driven 
transitions~$\vert 1\rangle_0\longleftrightarrow\vert 2\rangle_{-+}$ 
and~$\vert 1\rangle_0\longleftrightarrow\vert 2\rangle_{--}$. 
Finally, peak {\em i} vanishes if
we eliminate the~$\omega_2$--driven resonant 
pathway~$\vert 0\rangle\longleftrightarrow\vert 1\rangle_-$
followed by 
the~$\omega_1$--driven 
$\vert 1\rangle_-\longleftrightarrow\vert 2\rangle_{-+}$ transition 
(figures~\ref{fig:sparsebeam:es01}(e,~f) and~\ref{fig:sparsebeam:ef12}(e,~f)). 

The clearest 
unambiguous spectroscopic feature of the two 2LA--cavity system, which
cannot be accounted for by the (single--atom) JC--spectrum, is peak {\em i}. 
This peak is almost visible in figure~\ref{fig:smeared_gf63}(a)
and clearly visible after background subtraction as
shown in figure~\ref{fig:smeared_gf63}(b).
In addition, the peak is
much more pronounced than the other relevant 2PCR peaks at~{\em iv}, {\em v}
and {\em vii}. Thus, this peak provides an ideal 
signature of quantum field effects in two--atom cavity systems.  

Furthermore, the method of suppressed transitions
allows us to identify the 
deleterious two--photon pathways
that destroy the quantum signature of
single--atom and two--atom peaks
(single--atom peaks  {\em ii} and {\em iii} 
and two--atom peaks
{\em i}, {\em iv}, {\em v} and {\em vii}). These
pathways are responsible for the difference between 
figure~\ref{fig:smeared_gf63}(a)
and~\ref{fig:smeared_gf63}(b). Background subtraction eliminates precisely
those deleterious pathways. 
Figures~\ref{fig:sparsebeam:es12}(g)
and~\ref{fig:sparsebeam:es01}(e) clearly show that peaks {\em i}, {\em ii} 
and {\em iii}
are  strongly affected by the 
$\vert 0\rangle\longleftrightarrow\vert 1\rangle_-
\longleftrightarrow\vert 2\rangle_{-+}$ pathway via two~$\omega_2$ photons. 
The second major undesired pathway washes peaks {\em iv}, {\em v}, 
{\em vi} and {\em vii} away. 
This pathway is primarily due to 
the~$\omega_2$--driven transition $\vert 0\rangle\longleftrightarrow
\vert 1\rangle_+$
(as shown in figure~\ref{fig:sparsebeam:es01}(a)), followed 
by~an~$\omega_2$--driven transition
$\vert 1\rangle_+\longleftrightarrow \vert 2\rangle_{++}$ (not shown).

\section{Conclusion}
\label{sec:conclusions}

For an atomic beam the number of atoms in the cavity mode is a 
randomly varying 
quantity. For a sparse beam, most of the time there is effectively 
no atom in the cavity. A large portion of time there is effectively 
one atom, although, of course, its position is a randomly varying
quantity. There is also a 
contribution to the photon coincidence spectrum due to two or more atoms 
simultaneous interacting with the cavity mode. In 
section~\ref{sec:twoatomspec}, 
we have assumed precisely two atoms in the cavity but left
the positions random. In this section we include the signal for one
and no atom events in the cavity with appropriate weightings for a sparse
beam. We show that the 2PCR reveals genuine quantum field effects for
two--atom events.

\begin{figure}
\begin{picture}(200,430)(20,-220)
\font\gnuplot=cmr10 at 10pt
\gnuplot
\rotatebox{90}{\resizebox{400pt}{420pt}{\includegraphics{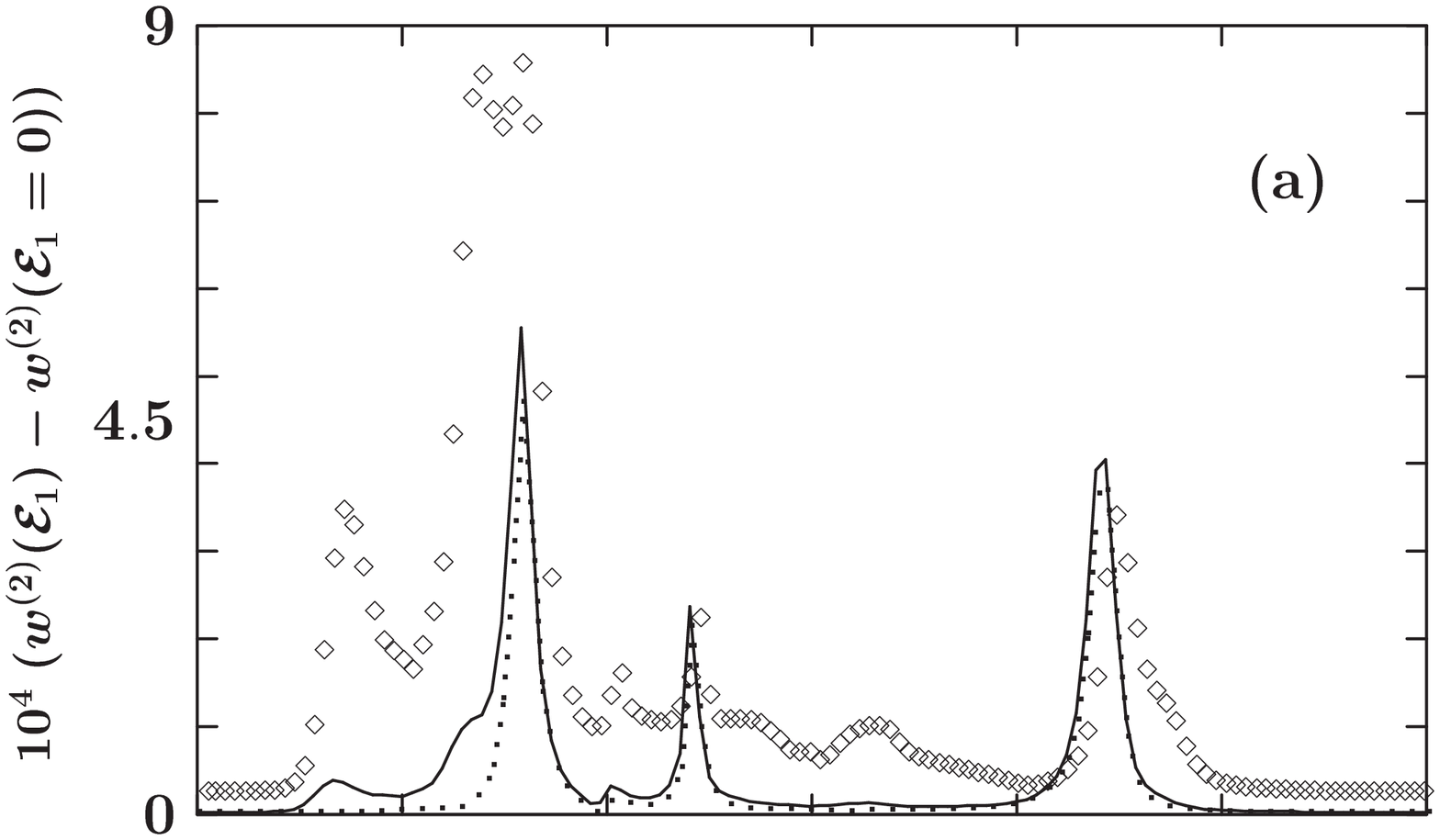}}}
\end{picture}
\begin{picture}(200,330)(220,20)
\font\gnuplot=cmr10 at 10pt
\gnuplot
\rotatebox{90}{\resizebox{400pt}{420pt}{\includegraphics{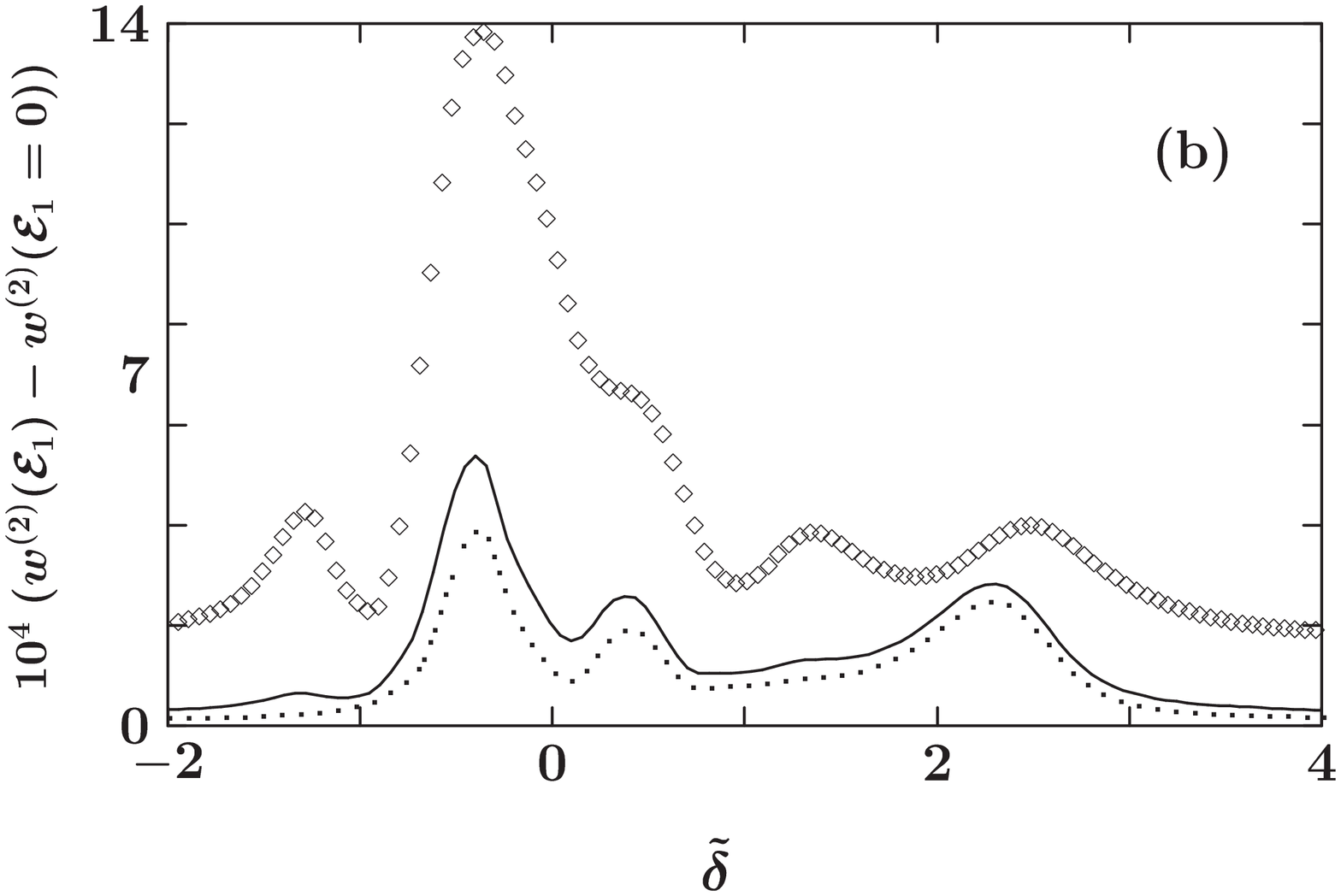}}}
\end{picture}
\caption{Two-photon count rate vs normalised 
	scanning 
	frequency for one atom (dotted line), two atoms ($\Diamond$) and
	a sparse atomic beam  (solid line) with~$p_1/p_2=9$ for 
	${\cal E}_1/\kappa=1/\sqrt2$,
	${\cal E}_2/\kappa=\sqrt{2}$ and $\gamma/\kappa=2$
	with (a)~$g_f=63\kappa$ and (b)~$g_f=9\kappa$.  
	}
\label{fig:sparsebeam} 
\end{figure}

We let~$\bar{\rho}_n$ 
represent
the density matrix for~$n$ atoms in the cavity, averaged over the atomic 
positions. This matrix~$\bar{\rho}_n$ is a generalisation of 
equation~(\ref{density:matrix}) where the subscript refers to the 
number of atoms in the cavity. 
For the sparse atomic beam, we will ignore $\bar{\rho}_n$ for 
$n>2$. The 
case~$n=0$ is not significant because off--axis driving ensures that photon 
coincidences do not arise when there is no atom present. Here we make the 
reasonable assumption that $p_1/p_2=9$. That is, the probability of
having one atom in the cavity is nine times more likely than having
two atoms. This is compatible with the sparse beam assumption. The
result for~$p_1/p_2=9$ is shown in figure~\ref{fig:sparsebeam}(a,~b). 

We consider two cases: a high value of $g_{f}\, ( =63\kappa )$ as
discussed in subsection~\ref{subsec:msuptrans} 
($g_{f}=63\kappa$ was assumed in a study of
multiphoton coincidence spectroscopy for the JC model~\cite{Hor99}), and a
low value of $g_{f}\,( =9\kappa )$, which has been the subject of other
studies of photon coincidence spectroscopy~\cite{Car96,San97,Hor00}.  The
case of a high value of~$g_{f}$ corresponds to strong coupling and is
depicted in figure~\ref{fig:sparsebeam}(a); 
the low value case is depicted in figure~\ref{fig:sparsebeam}(b).

For $g_{f} =63\kappa$, we observe that peak~$i$, which corresponds to
a genuine quantum field effect with two strongly coupled atoms in the
cavity, is still quite pronounced, despite the strong signal from 
single--atom effects. Even for $p_1/p_2=9$, the two--atom events contribute a 
strong difference--2PCR signal.  Therefore, we can assert that two--photon
coincidence spectroscopy is an effective means for extracting the quantum
field signature of two atoms in a quantised field cavity for a sparse atom
beam. 

On the other hand, we explore the case of relatively low coupling, namely
$g_{f}=9\kappa$. This case is depicted in figure~\ref{fig:sparsebeam}(b) 
and shows that
the single--atom 2PCR spectrum (depicted as a dotted line) is not modified
in any substantial way by the two--atom contributions.  Therefore, the
employment of 2PCS as a probe of quantum field effects in the
(single--atom) JC model is sound for a sparse atomic beam. Of course it
was assumed in references~\cite{Car96,San97} that multi--atom effects were not
significant factors in modifying the ideal single--atom two--photon
spectrum.  However, here figure~\ref{fig:sparsebeam}(b) reveals quite 
clearly how much of an
effect arises from two--atom events and how small this effect is.

On the one hand, for strong coupling, the two--atom effect is large, and
peak~$i$, in particular, can be used to probe experimentally the quantum
field effect for two--atom cavity quantum electrodynamics.  On the other
hand, for lower coupling strengths, we can safely ignore the effects of
rare two--atom events in cavity quantum electrodynamics.  The theory
presented here allows us to distinguish the two cases and know when
two--atom effects are important.

\section*{Acknowledgements}
We have benefitted from valuable
discussions with H.\ J.\ Carmichael. 
This research has been supported by Australian Research Council
Large and Small Grants, by an Australian Research Council
International Research Exchange (IREX) grant,  
by a Macquarie University Research Grant and by the Macquarie
University Postgraduate Research Fund. 

\appendix

\section{Eigenvectors and eigenvalues for two two--level atoms in
as single--mode cavity}
\label{app:a:eigenvec}

The Hamiltonian for two 2LAs in an optical cavity can be derived from
equation~(\ref{H:mult}) and is given by
\begin{eqnarray}
\label{eq:twoatom:H}
H &=&\hbar\omega a^{\dag} a+
\hbar\omega\sigma_z^{(1)}
+\hbar\omega\sigma_z^{(2)} 
+i\hbar g_1  \left( a^{\dag}\sigma_-^{(1)}-a\sigma_+^{(1)}\right)
\nonumber \\ 
&&+ i\hbar g_2  \left( a^{\dag}\sigma_-^{(2)}-a\sigma_+^{(2)}\right).
\end{eqnarray}
For the one--atom case, the Hamiltonian is diagonalised to
yield the JC dressed states. Diagonalising equation~(\ref{eq:twoatom:H})
for~$g_1=g_2$ yields the dressed states of the two--2LA Tavis--Cummings
system~\cite{Tav69}. 
In this section, we consider the dressed states for~$g_1\neq g_2$. 

\begin{figure}
\begin{picture}(200,300)(0,0)
\font\gnuplot=cmr10 at 10pt
\gnuplot
{\resizebox{180pt}{280pt}{\includegraphics{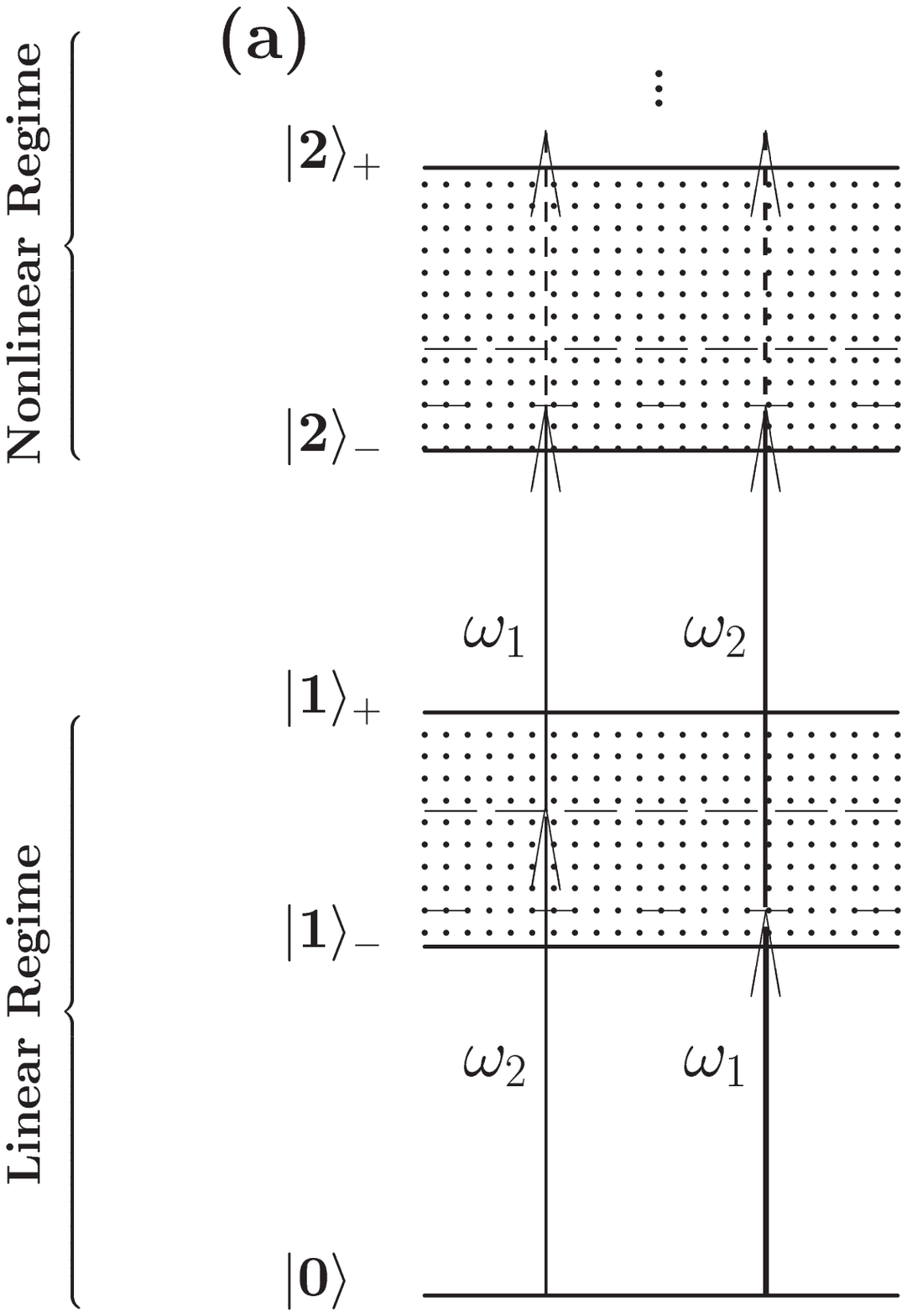}}}
\end{picture}
\begin{picture}(240,300)(-25,0)
\font\gnuplot=cmr10 at 10pt
\gnuplot
{\resizebox{145pt}{280pt}{\includegraphics{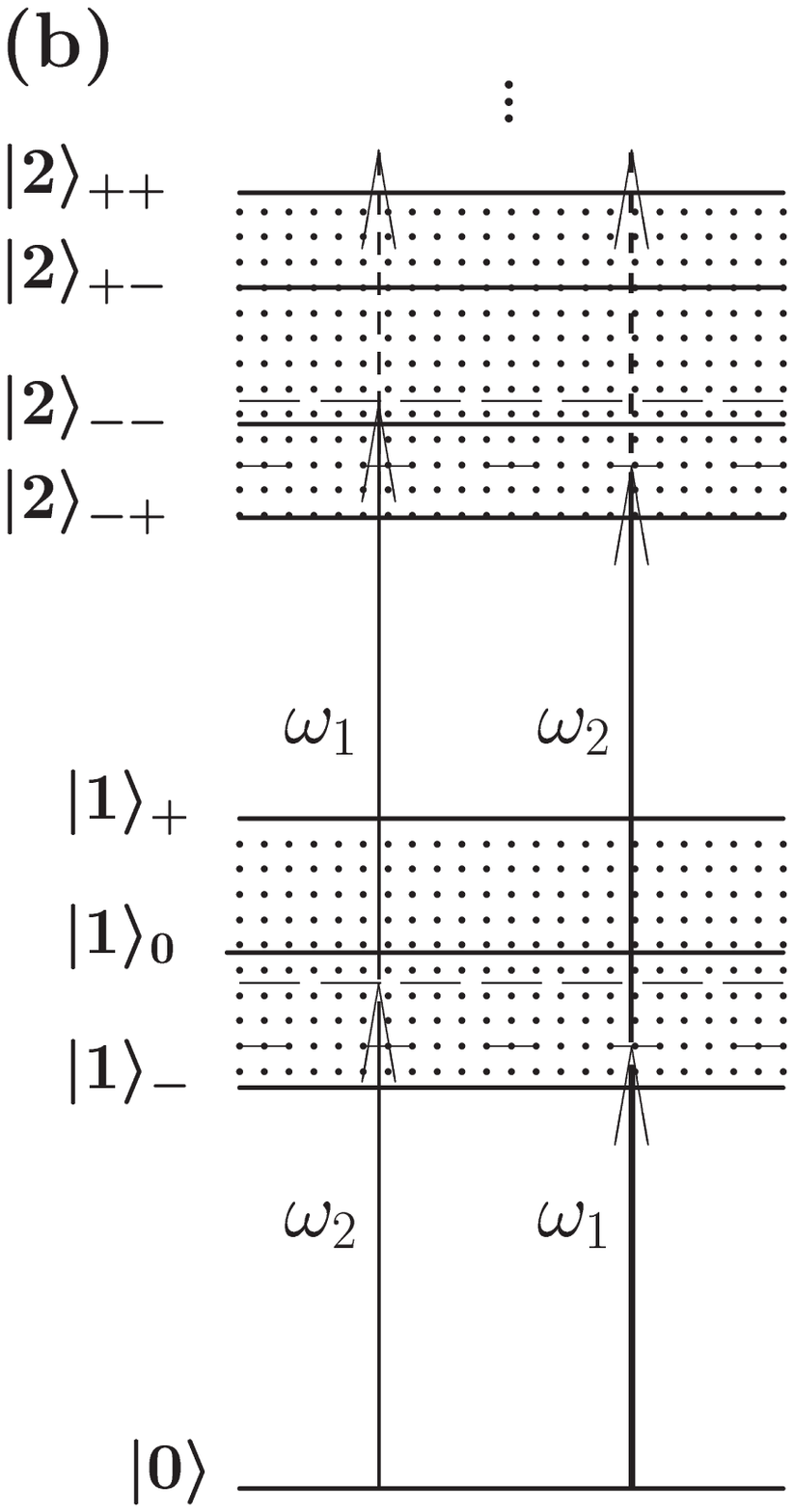}}}
\end{picture}
\caption{The lowest dressed--state multiplet in the nonlinear
regime for (a)~the Jaynes--Cummings system and (b)~two two--level atoms
coupled to a single quantised field mode of the cavity.
Each system may be driven by a bichromatic field with 
frequencies~$\omega_1$ and~$\omega_2$. Two excitation pathways to
the second multiplet are depicted in each case. Inhomogeneous broadening
of multiplets is depicted by the shaded region. In case (a),
the first doublet has inhomogeneous broadening~$2g_{\rm max}$, and the
second has width~$2\sqrt2 g_{\rm max}$. In case (b), the triplet has
an inhomogeneous broadening width of~$2\sqrt2 g_{\rm max}$. }
\label{fig:ladder}
\end{figure}

The (one--atom) JC 
spectrum of figure~\ref{fig:ladder}(a) is a ladder of states consisting
of a ground state and then a sequence of couplets. The dressed states are 
designated as~$\vert n \rangle_{\varepsilon}$, with~$n$ designating the
couplet number and~$\varepsilon\in\{+,\,-\}$ designating whether the  
state is the one with higher or lower energy. The
quantity~$n$ corresponds to the number of quanta in the system; for example the
states~$\vert 1\rangle_{\pm}$ are superpositions 
of~$\vert 1\rangle\vert {\tt g}\rangle$ 
and~$\vert 0\rangle\vert {\tt e}\rangle$,  i.e.\ a 
superposition of a photon (and the atom in the ground state) with
an atomic excitation (and the cavity mode in the ground state).
Here we generalise this approach to the case of two 2LAs in the cavity.

The first few levels of the spectral ladder are shown in 
figure~\ref{fig:ladder}(b). The ground state is designated as~$\vert 0\rangle$
and corresponds to an absence of photons in the cavity and both atoms in the 
ground state. For~$\tilde{g}^2\equiv g_1^2+g_2^2$, 
the set of one--quantum states form a triplet, and
the three eigenstates of this triplet are 
\begin{eqnarray}
\label{eq:1_0}
\vert 1\rangle_0 &=&\tilde{g}^{-1}\vert 0\rangle(g_2\vert {\tt e}\rangle
	\vert {\tt g} \rangle-g_1\vert {\tt g}\rangle\vert {\tt e}\rangle) \\
\label{eq:1_pm}
\vert 1\rangle_{\pm} &=& \mp 2^{-1/2} i\vert 0\rangle
	\vert {\tt g}\rangle \vert {\tt g}\rangle
	+2^{-1/2}\tilde{g}^{-1}\vert 0\rangle (
	g_2\vert {\tt g}\rangle\vert {\tt e}\rangle
	+ g_1\vert {\tt e} \rangle\vert {\tt g}\rangle),
\end{eqnarray}
with corresponding eigenvalues 
\begin{eqnarray}
\label{eq:pm}
\lambda^{\pm} &=& \omega\pm\tilde{g} \quad 
	\mbox{(for $\vert 1\rangle_{\pm}$)} \\
\label{eq:0}
\lambda^{0} &=& \omega \quad \quad \,\,\,\,\, \mbox{(for $\vert 1\rangle_0$)}, 
\end{eqnarray}
respectively.

The multi--quanta states, for two and more energy quanta, are
quadruplets. We use the notion~$\vert n\rangle_{\varepsilon\,\varepsilon'}$,
with~$n$ the number of quanta ($n=2$ for the first quadruplet, $n=3$
for the second, etc.). The subscripts~$\varepsilon$ 
and~$\varepsilon'$ are each either~$+$ or $-$ 
(i.e.~$\varepsilon,\, \varepsilon'\in \{+,\, -\}$), and there are 
four distinct combinations of~$\varepsilon\, \varepsilon'$
corresponding to each of the four states in 
the quadruplet. The choices of~$\varepsilon$ and~$\varepsilon'$ are
determined by signs in the expression for the eigenvalue of the
particular state. 
The four states of the~$n^{\rm th}$ quadruplet are
\begin{eqnarray}
\vert n+1 \rangle_{+\,\pm} \!\!\!&=& \!\!\!
		\Lambda_\pm^{(n)}\vert n+1\rangle 
				\vert {\tt g}\rangle \vert {\tt g}\rangle
		- i\zeta_{1\pm}^{(n)}\vert n\rangle\vert {\tt g}\rangle
				\vert {\tt e}\rangle \nonumber \\
		& & - i\zeta_{2\pm}^{(n)}\vert n\rangle\vert {\tt e}\rangle
				\vert {\tt g}\rangle
		+ ({\cal N}^{(n)}_\pm)^{-\frac{1}{2}}
		\vert n-1 \rangle\vert {\tt g}\rangle\vert {\tt g}\rangle
		\nonumber \\
\vert n+1 \rangle_{-\,\pm} \!\!\!&=& \!\!\!
		\Lambda_\pm^{(n)}\vert n+1\rangle 
				\vert {\tt g}\rangle \vert {\tt g}\rangle
		+ i\zeta_{1\pm}^{(n)}\vert n\rangle\vert {\tt g}\rangle
				\vert {\tt e}\rangle \nonumber \\
		& &+ i\zeta_{2\pm}^{(n)}\vert n\rangle\vert {\tt e}\rangle
				\vert {\tt g}\rangle
		+ ({\cal N}_\pm^{(n)})^{-\frac{1}{2}}
		\vert n-1 \rangle\vert {\tt g}\rangle\vert {\tt g}\rangle.
		\nonumber 
\end{eqnarray}
The coefficients employ the convenient terms
\begin{eqnarray}
\Xi^{(n)} &=& \sqrt{(2 n+1)^2 (g_1^2+g_2^2)^2
-4n(n+1)(g_1^2-g_2^2)^2},	\nonumber \\
{\Lambda_\pm^{(n)}}' &=& -\frac{(g_1^2+g_2^2\pm\Xi^{(n)})}{4g_1 g_2
\sqrt{ n(n+1)}},
\nonumber \\
{\zeta_{1\pm}^{(n)}}' &=& \frac{g_2^2+g_1^2 (1+4n)\pm\Xi^{(n)}}{2 g_2
\sqrt{2n((2n+1)(g_1^2+g_2^2)\pm\Xi^{(n)})}}, \nonumber \\
{\zeta_{2\pm}^{(n)}}' &=& \frac{g_1^2+g_2^2 (1+4n)\pm\Xi^{(n)}}{2 g_1
\sqrt{2n((2n+1)(g_1^2+g_2^2)\pm\Xi^{(n)})}},\nonumber 
\end{eqnarray}
with
\begin{eqnarray}
& & {\cal N}^{(n)}_{\pm} = {{\Lambda_\pm^{(n)}}'}^2 + {{\zeta_{1\pm}^{(n)}}'}^2
		+ {{\zeta_{2\pm}^{(n)}}'}^2 +1, \nonumber \\
\Lambda_\pm^{(n)} &=& \frac{{\Lambda_\pm^{(n)}}'}{\sqrt{{\cal N}_\pm^{(n)}}},
\; \; \;\;\;\;\;
\zeta_{1\pm}^{(n)} = \frac{{\zeta_{1\pm}^{(n)}}'}{\sqrt{{\cal N}_\pm^{(n)}}},
\; \; \;\;\;\;\;
\zeta_{2\pm}^{(n)} = \frac{{\zeta_{2\pm}^{(n)}}'}{\sqrt{{\cal N}_\pm^{(n)}}}.
\; \; \;\;\;\;\;
\nonumber 
\end{eqnarray}
The corresponding eigenvalues of the $n$ quadruplet are 
\begin{eqnarray}
\label{eq:quadrupeig++}
\lambda^{\pm\pm}_n \!\!\!&=& \!\!\!\! (n+1)
\omega\pm\sqrt{\frac{(2n+1)(g_1^2+g_2^2)\pm\Xi^{(n)}}{2}}
\; \mbox{(for $\vert n+1\rangle_{\pm\pm}$)},
\\
\label{eq:quadrupeig-+}
\lambda^{\mp\pm}_n \!\!\! &=& \!\!\!\! (n+1)
\omega\mp\sqrt{\frac{(2n+1)(g_1^2+g_2^2)\pm\Xi^{(n)}}{2}}
\; \mbox{(for $\vert n+1\rangle_{\mp\pm}$)}.
\end{eqnarray}

For this analysis, we have assumed that~$g_1\neq  g_2$. 
If~$g_1=g_2$, we obtain the Tavis--Cummings model~\cite{Tav69},
for which the ladder consists of a ground state singlet, 
a doublet for the one--quantum states (with an 
energy splitting~$\sqrt2g$) and then triplets for two or more quantum states.
As this case has been studied in depth, we do not analyse the 
Tavis--Cummings model here.

\newpage

\end{document}